\documentclass{emss-arxiv}

\address[blanc@ann.jussieu.fr]{Xavier Blanc, Laboratoire Jacques-Louis Lions (UMR CNRS 7598), Universit\'e Paris Diderot, Bâtiment Sophie Germain, 5 rue Thomas Mann, 75205 Paris Cedex 13, France}
\address[mathieu.lewin@math.cnrs.fr]{Mathieu Lewin, CNRS and CEREMADE (UMR CNRS 7534), University of Paris-Dauphine, Place de Lattre de Tassigny, 75775 Paris Cedex 16, France}

\received[revision date]{submission date}

\usepackage[T1]{fontenc}
\usepackage[latin1]{inputenc}
\usepackage{epsfig}
\usepackage{amsmath,amssymb,euscript}
\usepackage{pst-all}
\usepackage{geometry}
\usepackage{dsfont}

\def\XXint#1#2#3{{\setbox0=\hbox{$#1{#2#3}{\int}$}
    \vcenter{\hbox{$#2#3$}}\kern-.5\wd0}}

\def\longrightharpoonup{\DOTSB\relbar\joinrel\rightharpoonup}

\def\NN{\mathbb N}

\def\RR{\mathbb R}
\def\ZZ{\mathbb Z}

\newcommand{\ii}{\infty}
\newcommand\R{{\ensuremath {\mathbb R} }}
\newcommand\C{{\ensuremath {\mathbb C} }}

\newcommand\Z{{\ensuremath {\mathbb Z} }}

\newcommand\1{{\ensuremath {\mathds 1} }}
\newcommand{\cH}{\mathcal{H}}
\newcommand{\cE}{\mathcal{E}}
\newcommand{\cM}{\mathcal{M}}
\newcommand\nn{\nonumber}
\renewcommand\phi{\varphi}
\newcommand{\wto}{\rightharpoonup}

\renewcommand{\epsilon}{\varepsilon}
\def\and{\ and }

\theoremstyle{definition}

\title[The crystallization conjecture: a review]{The crystallization conjecture: a review}

\author[X. Blanc and M. Lewin]{Xavier Blanc  \and  Mathieu Lewin }

\begin{document}

\begin{abstract}
In this article we describe the \emph{crystallization conjecture}. It states that, in appropriate physical conditions,
interacting particles always place themselves into periodic configurations, breaking thereby the natural
translation-invariance of the system. This famous problem is still largely open. Mathematically, it amounts to
studying the minima of a real-valued function defined on $\R^{3N}$ where $N$ is the number of particles, which
tends to infinity. We review the existing literature and mention several related open problems, of which many have
not been thoroughly studied.

\medskip

\noindent \scriptsize Final version to appear in \emph{EMS Surv. Math. Sci.}.
\end{abstract}

\maketitle

\tableofcontents

\newpage

At the microscopic scale, most crystals are composed of atoms which are arranged on a periodic lattice. This specific geometric structure
has important consequences at the macroscopic scale. For instance, in snowflakes the atoms are arranged on an hexagonal lattice, which explains the beautiful
six-pointed figures that can be found in nature. The aim of crystallography is to study those periodic structures and their properties at larger scales.

2014 was declared the year of crystallography by UNESCO~\cite{Unesco-14} and this gives us the opportunity to draw attention to a difficult mathematical conjecture, also important from a physical point of view, which has been studied a lot without being completely solved. While crystallographers study the properties of some periodic arrangements and compare them, there remains a more fundamental question: why is it favorable (at low temperature) for the atoms to spontaneously arrange themselves on a periodic array? This periodic order seems to only appear in the limit of a large number of particles, which makes the question particularly difficult.

\medskip

In this article we rigorously formulate this long-standing problem and we make a review of the existing results as
well as of the remaining open questions. We will mostly discuss the simplest model (classical particles interacting
with a two-body potential at zero temperature), before addressing more advanced situations (for instance quantum
systems and/or positive temperature).

\section{The classical model}
\label{sec:classique}

\subsection{Energy}
\label{sec:interaction}

Let us consider a set of $N$ classical identical particles in $\R^d$ (in practice $d=1,2,3$), interacting by pairs through a potential $V$ depending only on the distance between them.
We denote by $x_1,\dots, x_N\in\RR^d$ and $p_1,\dots,p_N\in\RR^d$ the positions and momenta of these particles. The model to be used is that of the Hamiltonian dynamics, based on the energy
\begin{equation}
\cH_N \left(x_1,\dots,x_N,p_1,\dots, p_N\right) = \sum_{i=1}^N \frac{|p_i|^2}{2m} + \sum_{1\leq i < j\leq N} V\left(\left|x_i- x_j\right|\right).
 \label{eq:Hamiltonien}
\end{equation}
Here $m$ is the mass of the particles and $|\cdot|$ is the Euclidean norm of $\RR^d$.

At zero temperature, the equilibrium states are the minima of $\cH_N$, which all satisfy $p_1 = \dots = p_N = 0$. If one is only interested in those, it is therefore sufficient to consider the potential energy
$$\cE_N \left(x_1,\dots,x_N\right) = \sum_{1\leq i < j\leq N}
V\left(\left|x_i- x_j\right|\right),$$
and to understand how the $x_i$'s solving the minimization problem
\begin{equation}
  \label{eq:minimisation}
  E(N)= \inf \Bigl\{ \cE_N(x_1,\dots,x_N), \quad x_1,\dots,x_N\in \RR^d\Bigr\},
\end{equation}
are arranged in $\R^d$ in the limit $N\to+\infty$. Let us note that $\cE_N$ is invariant under translations and rotations. Any
configuration may be rotated and translated by a fixed vector without changing the total energy. Minimizers
of~\eqref{eq:minimisation} are thus not unique. At positive temperature one should consider the Gibbs measure
$\exp(-\cH_N/T)$, as will be discussed later in Section~\ref{sec:temperature} below.

In practice the potential $V$ depends on the type of atoms and is not explicitly known. As atoms are not elementary particles, $V$ cannot be deduced from first principles. It is therefore important to obtain mathematical results which are sufficiently generic with regards to $V$.

\begin{figure}
  \centering
  \includegraphics[width=7.2cm]{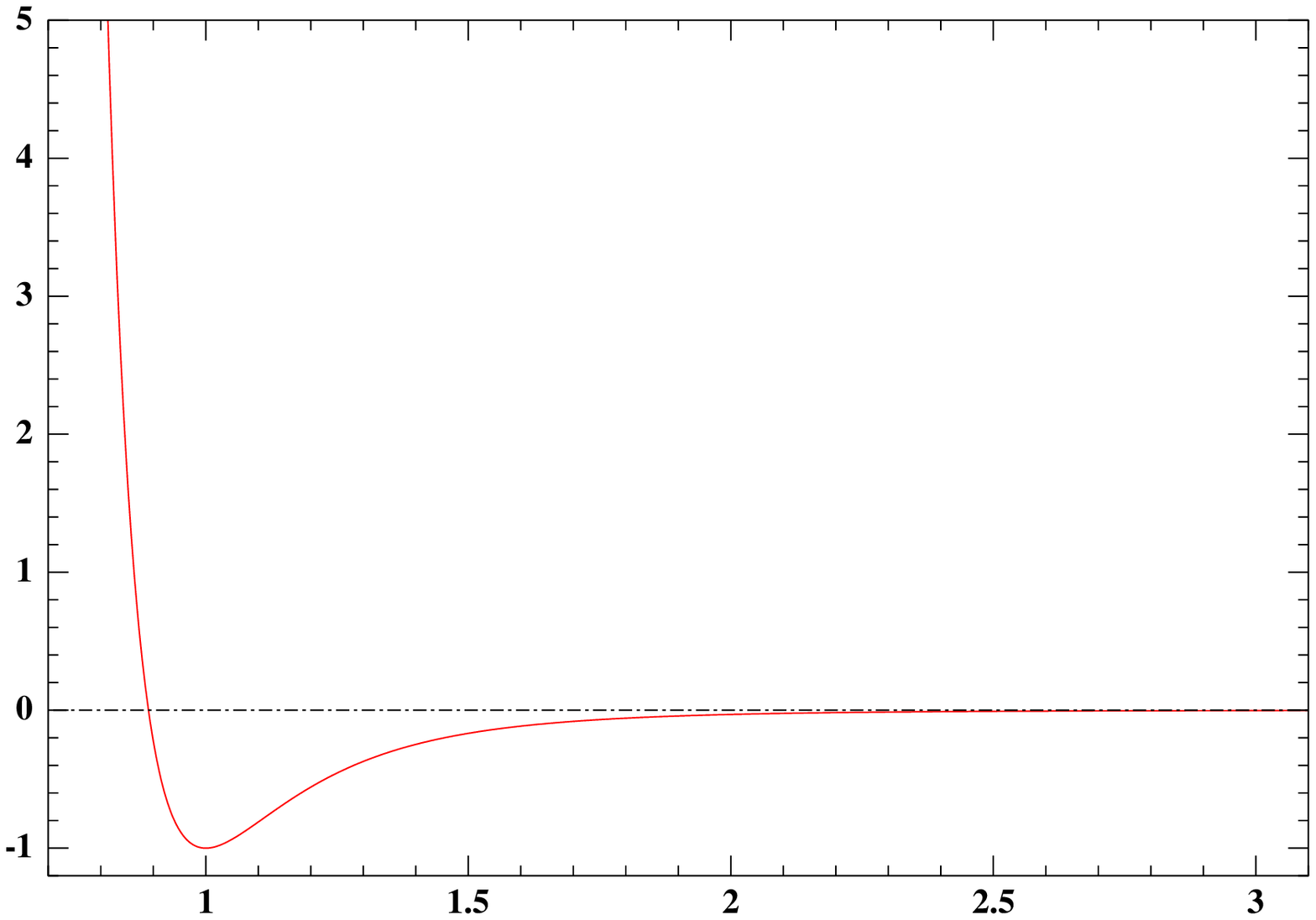}
  \includegraphics[width=7.2cm]{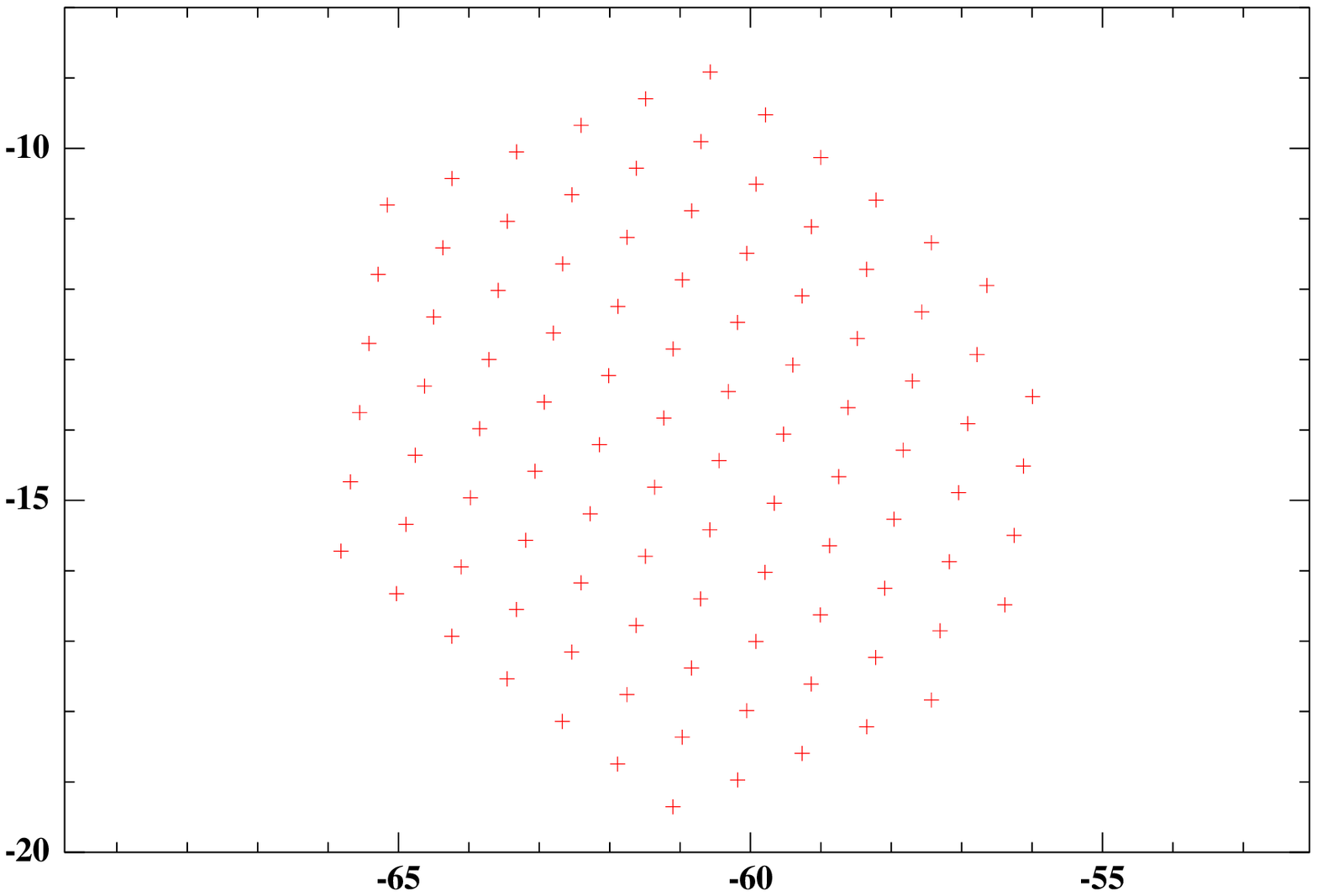}
  \caption{\small Left: the Lennard-Jones potential \eqref{eq:lj}. Right: a minimizer for the variational problem~\eqref{eq:minimisation}, computed numerically in~\cite{BenBlaBri-03}, with $N=100$ and $d=2$. The particles seem to arrange themselves on an hexagonal lattice, and to form a large cluster having the shape of an hexagon.
}
  \label{fig:lj}
\end{figure}

Qualitatively, the function $V$ is usually assumed to be positive (repulsive) at small distances and negative
(attractive) at large distances. Since the interaction between two atoms which are far from each other is small, we assume that $V(r)\to0$ as $r\to+\ii$. A typical and very popular example is the \emph{Lennard-Jones  potential}
\begin{equation}
  \label{eq:lj}
  V_{\rm LJ}(r) = \frac1{12}\left(\frac{r_0}{r}\right)^{12} - \frac16\left(\frac{r_0}{r}\right)^6,
\end{equation}
drawn in Figure~\ref{fig:lj}.
The behavior at infinity in $r^{-6}$ mimics the Van der Waals interaction, that is, the one for radially symmetric neutral particles. The behavior at $r=0$ is, on the other hand, completely empirical. The number $r_0>0$ is the equilibrium distance for two isolated particles. It may be seen from Figure~\ref{fig:lj} that for this specific potential in the plane, the solutions $x_i$ to the minimization problem~\eqref{eq:minimisation}
are approximately located on an hexagonal lattice and that they moreover form a big cluster having the shape of an hexagon. Nobody knows how to prove these observations rigorously and this is essentially the crystallization conjecture that will be discussed in this article.

For the rest of the article, we consider a general radial potential $V$ that tends to zero at infinity. Some assumptions are however necessary to ensure that our question is well posed.

\subsection{Binding: existence of minimizers for $E(N)$}
\label{sec:binding}

The first assumption that we need is that particles can bind, that is, $E(N)$ should possess at least one minimizer (maybe only for a subsequence $N_j\to\ii$).
For instance, if $V>0$, then
$E(N) = e = 0$ for any $N\in\NN$, but the minimization problem~(\ref{eq:minimisation}) has no solution. Indeed, the infimum of
$\cE_N(x_1,...,x_N)$ is reached only when the distances between particles $x_i$ tend to infinity. It is therefore mandatory to assume that $\min V\leq0$ (or to force the particles to stay together, as will be discussed in Section~\ref{sec:densite_fixe}). In physical situations the interaction is attractive at some distance and we therefore always have
$$\min V<0.$$
Note that $E(2)=\min V$, hence this implies that $E(2)<0$.

Next we remark that the energy is \emph{subadditive}, that is, it satisfies
\begin{equation}
  \label{eq:sous_additivite}
  \forall N, P\geq1, \quad E(N+P) \leq E(N)+ E(P).
\end{equation}
This inequality is shown by sending $P$ particles at infinity and using the fact that $V\to0$. In other words, we write
$$E(N+P)\leq \lim_{|\tau|\to\ii}\cE_{N+P}(x_1,...,x_N,y_1+\tau,y_P+\tau)=\cE_N(x_1,...,x_N)+\cE_P(y_1,...,y_P)$$
and get~\eqref{eq:sous_additivite} after optimizing with respect to the $x_j$'s and $y_j$'s. Applying~\eqref{eq:sous_additivite} inductively, we find that
$$E(N)\leq \lfloor N/2\rfloor E(2)\leq (N-1)\frac{E(2)}{2}.$$
The optimal energy $E(N)$ is negative and bounded above by a term which behaves linearly in the particle number $N$. Since we require that $\min V<0$, we deduce that
\begin{equation}
\frac{E(N)}{N}\leq \frac{E(2)}{2}\left(1-\frac1N\right)<0.
\label{eq:upper_bound}
\end{equation}

When the strict \emph{binding inequalities}
\begin{equation}
E(N)<E(K)+E(N-K)
\label{eq:strict_binding_inequality}
\end{equation}
hold for all $K=1,...,N-1$ and when $V$ is a continuous function on $(0,\ii)$, then $E(N)$ can be shown to possess at least one minimizer.\footnote{The idea is to consider a minimizing sequence and to study whether some particles escape. If all the $x_j$'s are uniformly bounded (after applying an appropriate translation) then we get the existence of a minimizer using the continuity of $V$. If $K$ particles escape, then we get a contradiction from~\eqref{eq:strict_binding_inequality}.}
The verification of~\eqref{eq:strict_binding_inequality} can be complicated for a general $V$, but it is easy if $V<0$ at infinity. In this case,~\eqref{eq:strict_binding_inequality} can be proved by induction on $N$, using that if two groups of particles are far away, they always attract each other. We conclude from this discussion that when $V$ is continuous on $(0,\ii)$ and negative at infinity, then $E(N)$ always possesses minimizers and satisfies~\eqref{eq:upper_bound}.

It can also be proved that $\frac{E(N)}{N(N-1)}$ is non-decreasing, which gives us an inequality in the reverse order:
\begin{equation}
\frac{E(N)}{N}\geq (N-1)\frac{E(2)}{2}.
\label{eq:bad_lower_bound}
\end{equation}
The latter is indeed obvious from the formula of $\cE_N$, since each of the $N(N-1)/2$ terms in the sum can be bounded from below by $V(|x_i-x_j|)\geq \min(V)=E(2)$. As we will explain in the next section, the lower bound~\eqref{eq:bad_lower_bound} is not optimal in physically interesting cases.

\subsection{Stability and the behavior of $E(N)$ for large $N$}
\label{sec:stabilite}

We have seen that $E(N)$ is bounded above by a linear term in $N$. This linear behavior is indeed the interesting physical case and we will always require that the following limit
\begin{equation}
  \label{eq:limite_thermo}\boxed{
e_\ii = \lim_{N\to+\infty} \frac{E(N)}{N},}
\end{equation}
exists and is finite. The reason is the following: if we gather two macroscopic identical systems (for
a ``real life'' object, $N\approx 10^{23}$), the formation energy is by definition equal to
to $2E(N)-E(2N)>0$, which may be arbitrarily large if $E(N)$ is super-linear.\footnote{For instance $2 E(N)-E(2N)\sim C(2^a-2)N^a$ if $E(N)\sim-CN^a$ with $a>1$ and $C>0$.}
Note that $e_\ii<0$ from~\eqref{eq:upper_bound}, since we require that $\min(V)<0$.

It is clear that the existence of the limit~\eqref{eq:limite_thermo} implies that there is a lower bound of the form
\begin{equation}
  \label{eq:stabilite}
  E(N) \geq -C N,
\end{equation}
but the converse assertion is actually true and it is sometimes called Fekete's subadditive lemma~\cite{Fekete-23}. Using that $E(N)\leq \lfloor N/K\rfloor E(K)$ for every fixed $K$ and taking the limit $N\to\ii$, we even deduce
$$e_\ii=\inf_{N\geq1}\frac{E(N)}{N}.$$
In other words, $C=-e_\ii$ is the optimal constant in~\eqref{eq:stabilite}.
So we have to restrict ourselves to the potentials $V$ for which~\eqref{eq:stabilite} is satisfied, and those are called~\emph{stable} in the literature. The lower bound~\eqref{eq:stabilite} can be rewritten in the form
\begin{equation}
  \label{eq:stabilite2}
  \sum_{1\leq i<j\leq N} V(|x_i-x_j|) \geq -C N
\end{equation}
for all $N$ and all $x_1,...,x_N\in\R^d$. Stable potentials have been widely studied since the 60s
\cite{Ruelle-63a,Ruelle-63b,Dobrushin-64,Fisher-64,FisRue-66,Ruelle,Lanford-73}.

The
simplest example of a stable potential is
\begin{equation}
V = V_1 + V_2, \quad\text{with}\quad V_1\geq 0,\quad \widehat{V_2}\geq 0\quad
\text{and}\quad \int_{\RR^d} \widehat{V_2} <+\infty,
\label{eq:typical_stable_potential}
\end{equation}
where $\widehat{V_2}$ denotes the Fourier transform of $x\mapsto
V_2(|x|)$, cf.~\cite[Prop. 3.2.7]{Ruelle}. For $V_2$, the proof relies on the observation that
\begin{align}
\sum_{1\leq i<j\leq N}V_2(|x_i-x_j|)&=\frac12\sum_{i=1}^N\sum_{j=1}^NV_2(|x_i-x_j|)-\frac{N}{2}V(0)\nn\\
&=\frac1{2(2\pi)^{d/2}}\int_{\R^d}\widehat{V_2}(k)\Big|\sum_{j=1}^Ne^{ik\cdot x_j}\Big|^2\,dk-\frac{N}{2}V(0) \geq -\frac{N}{2}V(0).\label{eq:argument_positive_Fourier}
\end{align}
However, there are many physical potentials that cannot be written in the form~\eqref{eq:typical_stable_potential}.

Another famous example is that of a potential which is non-integrable at 0 but is bounded from below by an integrable function at infinity. Namely, $V$ is stable when
$$V(r)\geq
\begin{cases}
\phi_1(r) & \text{for $0\leq r\leq a$}\\
-C & \text{for $a\leq r\leq b$}\\
-\phi_2(r) & \text{for $r\geq b$}
\end{cases}$$
with $\phi_1$ and $\phi_2$ two positive decreasing functions on $(0,a)$ and $(b,\ii)$ respectively, such that
$$\int_0^a\phi_1(r)\,r^{d-1}\,dr=\ii,\qquad \int_b^{\ii}\phi_2(r)\,r^{d-1}\,dr<\ii,$$
see~\cite{Dobrushin-64} and~\cite[App.~A]{FisRue-66}. The idea behind these conditions is that the fast increase at zero prevents the particles from being too close to each other in average. The integrability of $V$ at infinity then makes the double sum in $\cE_N$ behave like $N$. The Lennard-Jones potential~\eqref{eq:lj} satisfies these conditions in dimensions $d\leq 5$, and it is therefore stable.

\subsection{Formation of a macroscopic object}
\label{sec:macroscopic}

The assumptions~(\ref{eq:sous_additivite})--(\ref{eq:stabilite}) above imply the
existence of a thermodynamic limit (\ref{eq:limite_thermo}), but they do not
ensure that a macroscopic object is formed in this limit. Their aim is
actually to avoid a collapse of the system by preventing particles to be
too close to each other. It is still possible that, in the optimal
configuration, particles do not stay close. This situation
should not be allowed.

In order to have the formation of a macroscopic object, we want that the
minimizing configuration of $N$ particles fill a volume of size $N$, in
the limit $N\to+\infty$. Moreover, the particles should be ``evenly spaced''
in this volume, as it is clear in the example of the Lennard-Jones
potential shown in Figure~\ref{fig:lj}.

The mathematical formulation of this property is not unique. One possibility is to apply a dilation of factor $N^{-1/d}$ to
the optimal configuration (this is a way to pass to the macroscopic
scale), and to consider the empirical measure
\begin{equation}
{\rm M}_N = \frac 1 N \sum_{i=1}^N \delta_{\frac{x_i}{N^{1/d}}}.
\label{eq:def_mu_dilatee}
\end{equation}
We ask if, after extracting a subsequence $N_j$ and translating the whole system by a vector $\tau_j$,
\begin{multline}
  \label{eq:cv_mu}
  {\rm M}_{N_j}(\cdot-\tau_j) \longrightharpoonup {\rm M} \text{ weakly-$\ast$ in the sense of measures, with ${\rm M}\in L^\infty(\RR^d)$,}\\ \text{$\operatorname{supp}({\rm M})$  compact and } \int_{\RR^d}{\rm M}(x)\,dx = 1.
\end{multline}
This means that the macroscopic object is included in the support of
${\rm M}$, and that, at this scale the system is continuous, with the
function $\rm M$ as local density. If~\eqref{eq:cv_mu} is satisfied, a
macroscopic object has formed. The knowledge about the positions of
particles is very crude, because of the dilation of factor $N^{-1/d}$,
which does not account for the local behavior of the system.

Finding conditions on the potential which imply the existence of the
weak limit \eqref{eq:cv_mu} is an important problem. However, it has
never been, to our knowledge, studied mathematically. For the
Lennard-Jones potential $V_{\rm LJ}$ in dimension $2$,
Figure~\ref{fig:lj} indicates that the limit ${\rm M}$ is proportional
to the characteristic function of a hexagon. In this example, the shape
of the support of $\rm M$, which is visible at the macroscopic scale (as
for instance snow flake structure), is a manifestation of the
crystalline order at the microscopic scale.

In the next section we discuss the crystallization conjecture, which concerns the microscopic properties of the system. We will come back to the macroscopic scale later in Section~\ref{sec:mu} below.

\section{The crystallization conjecture}\label{sec:conjecture}

\subsection{Formulation}
\label{sec:formulation}

\begin{figure}[h]
\centering
\includegraphics[width=6cm]{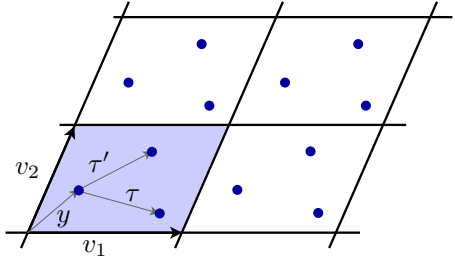}
\caption{\small Example of a periodic configuration in 2D.\label{fig:cristal}}
\end{figure}

We now come to the question which has been intensively studied since the
1960s, without being solved~\cite{Uhlenbeck-68,Radin-87}. This question is: does the system become
periodic in the limit $N\to\infty$? This may be formulated as follows.
Let us denote by
\begin{equation}
\mu_N=\sum_{i=1}^N\delta_{x_i}
\label{eq:def_mu_N}
\end{equation}
the empirical measure associated with the solution $x_1,\dots,x_N$ of
problem \eqref{eq:minimisation}. Note that, contrary to
\eqref{eq:def_mu_dilatee}, we do not use any dilation, and this means that
we study the system at the microscopic scale. We ask if, after extracting a subsequence $N_j$ and translating the whole system by a vector $\tau_j$,
\begin{equation}
\mu_{N_j}(\cdot-\tau_j)\wto \mu
\label{eq:cv_mu_cristal}
\end{equation}
locally, where $\mu$ is a locally finite measure. We say that
crystallization occurs if $\mu$ is periodic, that is, if there exists a
discrete subgroup
\begin{equation}
G=\left\{\sum_{j=1}^d n_j v_j,\ n_j\in\Z\right\}\subset \R^d,
 \label{eq:groupe_G}
\end{equation}
generated by $d$ independent vectors $v_1,...,v_d\in\R^d$, such that
$\mu(\cdot+g)=\mu$ for all $g\in G$. In order to avoid trivial cases, we
assume here that $G$ is the maximal group satisfying this property. Put
differently, the period is supposed to be minimal. The invariance under
the action of $G$ does not imply that the particles are located on the
vertices of a periodic lattice. This would correspond to the stronger
hypothesis
\begin{equation}
\mu=\sum_{g\in G}\delta_{g+y},
 \label{eq:mu_reseau_parfait}
\end{equation}
for some fixed vector $y\in \RR^d$, defining the position of the lattice
in space. For instance, it is possible to have 3 particles in the unit
cell of the lattice $G$, which are repeated periodically, as in
Figure~\ref{fig:cristal}. In such a case, the configuration of particles
is the superposition of 3 shifted crystalline lattices, and the measure
has the form
$$\mu=\sum_{g\in G}\delta_{g+y}+\sum_{g\in G}\delta_{g+y+\tau}+\sum_{g\in G}\delta_{g+y+\tau'}.$$

\begin{figure}[h]
\centering
\begin{tabular}{ccc}
\includegraphics[width=3cm]{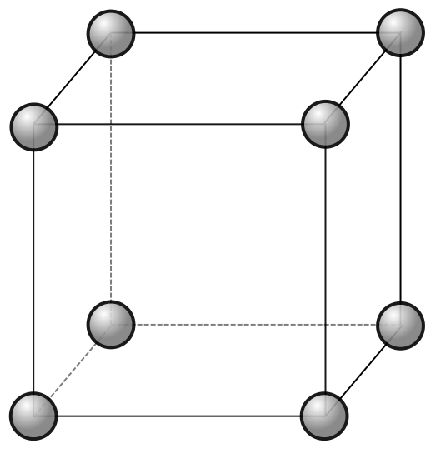}& \includegraphics[width=3cm]{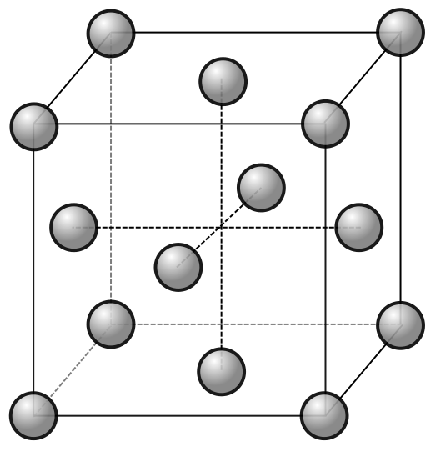} &\includegraphics[width=3cm]{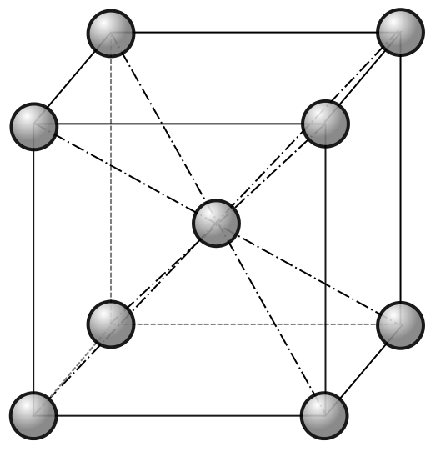} \\
\scriptsize simple cubic (SC) & \scriptsize face centered cubic (FCC) & \scriptsize body centered cubic (BCC)
\end{tabular}

\medskip

\begin{tabular}{c}
\includegraphics[width=8cm]{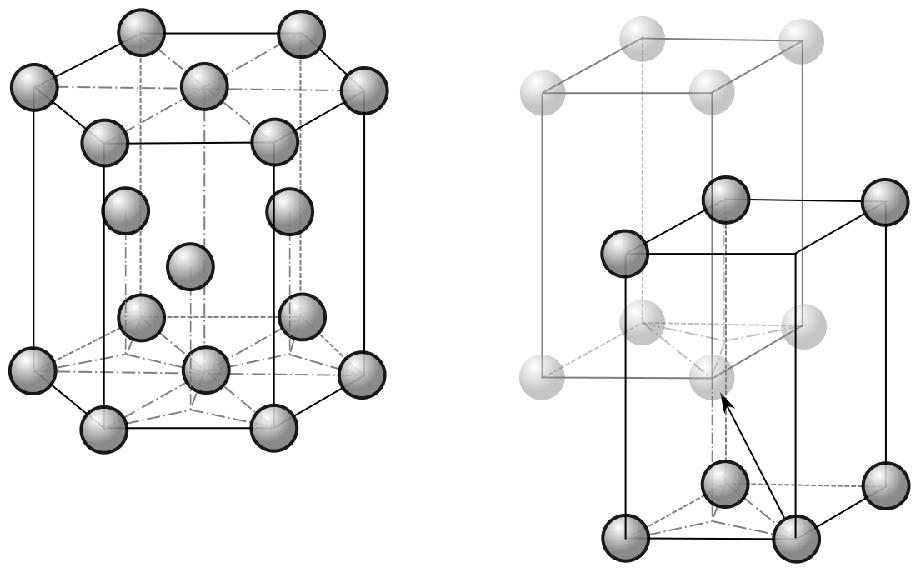}\\
 \scriptsize hexagonal close packed (HCP)
\end{tabular}

\caption{\small Most common configurations in 3D. \scriptsize\copyright~Wikipedia, GNU FDL, by~C.~Dang Ngoc Chan.\label{fig:reseaux}}
\end{figure}

In the special case where the particles are exactly on the nodes of the
lattice $G$, as in \eqref{eq:mu_reseau_parfait}, we use the word
\emph{Bravais lattice} or \emph{mono-atomic lattice}. For instance, in
dimension 3, the \emph{simple cubic lattice} (SC), \emph{face-centered
  cubic lattice} (FCC) and  \emph{body-centered cubic lattice} (BCC) are
all Bravais lattices. On the other hand, the \emph{hexagonal close packed
  lattice}~(HCP) is not. It is the superposition of two shifted Bravais
lattices (Figure~\ref{fig:reseaux}). This configuration is
the one used to pile up oranges in markets.

Crystallization may be seen as a symmetry breaking of the system: the invariance of the system under affine
isometries is lost in the process. If the positions of the particles form a periodic lattice, then applying a translation, rotation or reflexion to the system does not
change its energy. Hence, the set of minimizing lattices has the structure of the compact group
$\left(\RR^d/G\right) \ltimes O_d(\RR)$. Choosing a special minimizer for the positions $x_i$ at finite $N$, it is possible to select one of the limiting
lattices.

The microscopic scale convergence~\eqref{eq:cv_mu_cristal} does not give any information, in principle, about the
behavior at the macroscopic scale, such as the convergence of the dilated measure ${\rm M}_N$ defined
by~\eqref{eq:cv_mu}. Conversely, the convergence of ${\rm M}_N$ does not give any clue about that of
$\mu_N$. However, one
actually expects that the two phenomena are related. The understanding of the link between these two scales is
still incomplete, as we will discuss in Section~\ref{sec:mu} below.

Should crystallization be proved, the next question is to know which periodic configurations are present in the
limit (that is, what is the group $G$). Another question is to know if $\mu$ has the particular
form~\eqref{eq:mu_reseau_parfait} corresponding to a Bravais lattice. If not, one would ask how many particles
are present in each periodic cell, and what are their positions. In physical systems, lattices with larger symmetry
groups seem to be more common~\cite{Kittel}. These lattices are the hexagonal and square lattices in 2D, and the lattices presented in Figure~\ref{fig:reseaux} in 3D.

The ubiquity of crystals (at low temperature) indicates that crystallization is a universal phenomenon, which
should occur for a wide class of interaction potentials $V$. As we will see, several mathematical works prove
crystallization, but they are based on restrictive assumptions on $V$. To date, no generic class of potentials has
been identified, for which crystallization can be proved.

\subsection{Convergence of $\mu_N$ and the minimal distance between the particles}\label{sec:CV_mu_N}
Before discussing existing results about the crystallization problem in itself, we discuss when the sequence $\mu_N$ can be shown to have subsequences that converge locally, weakly-$\ast$ in the sense of measures, as we required in~\eqref{eq:cv_mu_cristal}. We need to show that the number of particles in an arbitrary fixed ball (centered at $\tau_N$ and of radius $R$) is uniformly bounded in the limit $N\to\ii$:
\begin{equation}
\#\{|x_j-\tau_N|\leq R\}\leq C(R)<\ii.
\label{eq:uniform_local_bound}
\end{equation}
If this property is satisfied, by usual compactness arguments we can then construct a subsequence of $\mu_N(\cdot-\tau_N)$ which converges locally to some $\mu$. We expect that $C(R)$ will behave like the volume $|B_R|$ of the ball for large $R$, but this is not needed at this point.

The condition~\eqref{eq:uniform_local_bound} immediately follows with $C(R)=|B_R|/|B_\epsilon|+O(R^{d-1})$ if we can prove that the smallest distance between the particles does not tend to zero in the limit $N\to\ii$:
$$\min_{1\leq i\neq j\leq N}|x_i-x_j|\geq 2\epsilon>0$$
for a minimizer of $\cE_N$. The idea is often that a very fast blow up of $V$ at zero should prevent that the particles get too close, but proving this very natural property can actually be a difficult task. It has been the object of several works: the Lennard-Jones potential
was covered in~\cite{XueMaiRos-92,MarFlo-92,Xue-97,Blanc-04,SchAddBomSch-07,Yuhjtman-15} and more general potentials were addressed in~\cite{LocSch-02,Vinko-05,VinNeu-07,AddSch-09}. As we will see in Section~\ref{sec:temperature}, the situation is much easier at positive temperature where it is sufficient to estimate the probability that some particles get very close.

\subsection{Crystallization results and sphere packing}\label{sec:empilement_compact}

\subsubsection*{In dimension $d=1$}

In dimension one, the problem of crystallization is rather well understood. The first results are due to Ventevogel
and Nijboer~\cite{Ventevogel-78,VenNij-79-1,VenNij-79-2}: they prove that the limit $e_\ii$ is reached by equidistant
configurations. This property is proved for a wide class of potentials $V$ (they are assumed to be non-increasing up
to a distance $r_0>0$, and non-decreasing for $r>r_0$, with additional hypotheses on $V''$), which includes the
Lennard-Jones potential $V_{\rm LJ}$. The convergence~\eqref{eq:cv_mu_cristal} is not proved in these works and is still
an open problem. It has been proved in the special case of $V_{\rm LJ}$ by Gardner and Radin \cite{GarRad-79}.

For some explicit examples of potentials $V$ (non-increasing up to a distance $r_0>0$, and non-decreasing for
$r>r_0$), it has been proved that the optimal configuration does not converge to a Bravais lattice. The limit can be
clusters of particles which are globally periodic~\cite{Ventevogel-78}. With an oscillating potential $V$, it is
even possible to obtain configurations which have no periodicity~\cite{HamRad-79}. In the latter, it is also
proved that such aperiodic configurations can be found as minimizers of a potential $V$ which is an arbitrarily
small perturbation of a potential for which crystallization occurs. This indicates that crystallization is an
unstable property if $V$ is perturbed by a small but highly oscillating function.
Thus, the conditions on $V$ ensuring crystallization are probably complex and have not been
completely understood yet, even in one dimension. It is commonly assumed that the interaction potential is smooth,
stable, non-increasing up to a distance $r_0$, and non-decreasing for $r>r_0$. However, no crystallization result has
been proved under these assumptions only, even in one dimension.

\subsubsection*{In dimension $d\geq2$}

\begin{figure}[h]
\centering
\includegraphics[width=8cm]{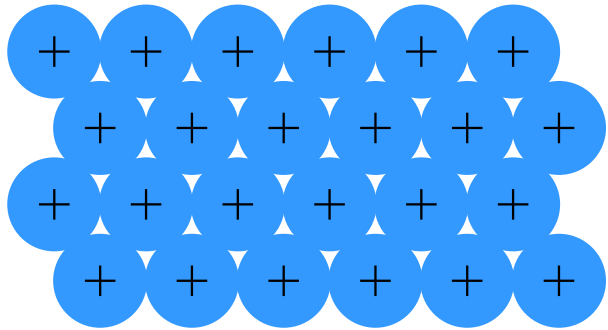}
\caption{\small Packing of identical disks, maximizing the density. The centers of the disks lie on a hexagonal lattice.\label{fig:hexagonal}}
\end{figure}

In higher dimensions, the problem is far from being understood. Most results are based on geometrical arguments,
which allow to reduce the question to the sphere packing problem. This question consists in finding the position of
non-overlapping spheres of equal radii giving the largest possible density. In two dimensions, the solution is
precisely the hexagonal lattice (see Figure~\ref{fig:hexagonal}). Thue has given two proofs of this result (in 1892
and in 1910), which both happened to contain flaws. A correct proof was then provided in 1940 by T\'oth
\cite{Rogers-64,ConSlo-99}. In dimension three, the problem is significantly more difficult. Kepler formulated it
in 1611, and it is therefore often called \emph{Kepler's conjecture}. A computer-assisted proof was given by Hales in 1999, then
published in 2005 in \cite{Hales-05}. Only recently (August 2014) has it been fully validated, after eleven years of
work by the \emph{Flyspeck} team~\cite{Flyspeck}, who managed to give a formal proof based on the softwares
\emph{Isabell} and \emph{HOL Light}. An important difference with the two-dimensional case is that, in 3D, the
problem has many solutions, including the hexagonal close packed lattice, the face centered cubic lattice and even non-periodic arrangements. The fact that FCC is
the unique minimizer among Bravais lattices was proved by Gauss~\cite{Gauss-40}.

The link between the crystallization problem and the sphere packing problem has been highlighted by Heitmann and
Radin in~\cite{HeiRad-80}. Indeed, if the interaction potential $V$ is given by
\begin{equation}
  \label{eq:sticky-disks}
  V(r) =
  \begin{cases}
    +\infty & \text{if } 0\leq r <1, \\
    -1 & \text{if } r=1, \\
    0 & \text{if } r>1,
  \end{cases}
\end{equation}
then the particles can be considered as hard spheres of radius $1/2$. These spheres tend to touch due to the condition $V(1)=-1$.
The crystallization problem is thus equivalent to the sphere packing, and one
obtains that the solution is the hexagonal lattice in 2D, and either FCC or the other sphere packing solutions in 3D.

Subsequent works aimed at generalizing this result to potentials which are similar to \eqref{eq:sticky-disks}, but
are closer to physically realistic interactions. For instance, in  \cite{Radin-81}, Radin considered a potential
satisfying \eqref{eq:sticky-disks} for $r\in [0,1]$, which is non-decreasing for $r\geq 1,$ and tends to $0$ fast
enough as $r\to +\infty$. In a famous article \cite{Theil-06}, Theil dealt with smoother, more realistic
potentials (which look like $V_{\rm LJ}$), in dimension two. However, he still used restrictive hypotheses on $V$.
This work has been extended to dimension three recently in \cite{FlaThe-13}, in which an additional
three-body term is added, which favors particular angles between bonds. A similar strategy has been used in
dimension $d=2$ in~\cite{ELi-09,MaiPioSte-14,MaiSte-14}, where the optimal lattice may be a square lattice. One can
therefore consider that the problem is not completely understood in dimension two, and completely open in
dimension three.

All these results in dimensions two and three rely heavily on the similarity with the sphere packing
problem. However, it is not clear if this should be the correct physical explanation. This would
exclude, for instance, configurations which are periodic but not mono-atomic. In such a case, particles form small
groups which are repeated periodically. The crystallization conjecture for a more general class of potentials is
still an open problem.

The sphere packing problem becomes more complex as the dimension increases. Note however that, in dimensions $d=4,8,24$, there are special lattices which are believed to solve the best packing problem~\cite{CohElk-03}. Although the
sphere packing problem in high dimension plays an important role in information theory, it is natural
to restrict ourselves to the (physically relevant) cases of dimension $d=1,2,3$. Indeed it has been conjectured that
crystallization  only occurs in small space dimensions for physical interactions~\cite{SkoDonStiTor-06,TorSti-06,TorStil-06b}.

\subsection{A variant: minimization at fixed density}\label{sec:densite_fixe}

It is possible to consider a potential $V$ which does not allow for the formation of a macroscopic object, if we force the particles to stay together. The idea is to
minimize the energy while keeping the density of particles $\rho$ fixed. This may be done by confining the particles in a large domain $\Omega$ and imposing that their number be $N\simeq \rho|\Omega|$, where $|\Omega|$ is
the volume of $\Omega$. This way, we get a family of problems depending on the parameter $\rho$.

To be more precise, we consider the minimization problem for $N$ particles in the domain $\Omega$
\begin{equation}
E_{\Omega}(N)=\inf \Bigl\{ \cE_N(x_1,\dots,x_N), \quad x_1,\dots,x_N\in \Omega\Bigr\},
\label{eq:min_rho_fixee}
\end{equation}
and we study the limit
\begin{equation}
\boxed{e(\rho)=\lim_{\substack{N\to\ii\\ |\Omega_N|\to\ii\\ N/|\Omega_N|\to\rho}}\frac{E_{\Omega_N}(N)}{N}}
\label{eq:limite_rho_fixee}
 \end{equation}
where $\rho>0$ is fixed and $\Omega_N$ is a sequence of domains which covers the whole space in the limit
$N\to\ii$. This limit should not depend on the chosen sequence. For
instance, it is often assumed that the measure of the boundary of $\Omega_N$ is a lower order term compared to its
volume $|\Omega_N|$~\cite{Ruelle}. To fix the ideas, one can think of $\Omega_N$ as a cube of side length
$(N/\rho)^{1/d}$, or a convex symmetric domain of unit volume, dilated by a factor $(N/\rho)^{1/d}$, as for
instance a ball of radius proportional to $(N/\rho)^{1/d}$.

Since $E_\Omega(N)\geq E(N)\geq e_\ii\,N$, it is clear that $e(\rho)\geq e_\ii$ for all $\rho>0$, where $e_\ii$ is the constant defined in~\eqref{eq:limite_thermo}.
In order for the limit~\eqref{eq:limite_rho_fixee} to exist, we also need an upper bound. Since the particles are confined to the domain $\Omega$, we cannot send them to infinity anymore, hence we loose most of the properties of Section~\ref{sec:binding}. The energy is not necessarily subadditive as in~\eqref{eq:sous_additivite} and we may well have $E(N)>0$. Furthermore, if the function $V$ tends to zero too slowly at infinity, each of the particles in the domain $\Omega$ interacts with many of the other particles and the energy might grow faster than $N$. A natural condition is to assume that $V$ is integrable at infinity:
$$\int_b^\ii|V(r)|\,r^{d-1}\,dr<\ii.$$
We can then easily find a position of the particles that will give an energy of order $N$, hence $E(N)\leq CN$. Under this assumption and the usual stability condition~\eqref{eq:stabilite}, the limit in~\eqref{eq:limite_rho_fixee} can be shown to exist and to be independent of the sequence $\Omega_N$. Potentials that are not integrable are also sometimes considered, but then one should divide the energy by the appropriate power of $N$.

Whether the energy behaves linearly or not, the problem is to study the behavior of the particle positions $x_1,...,x_N$ solution to the minimization
problem~\eqref{eq:min_rho_fixee}, and the questions are similar to the preceding case. A difference is that the model
is no more invariant under affine isometries. Different extraction of the sequence of minimizers may in principle
give limiting lattices with different positions. In practice, the position of the limiting lattice is often
determined by the choice of a particular sequence $\Omega_N$ (see~\cite{Kunz-74} for a discussion of this aspect in
dimension $d=1$).

Since a cube $\Omega_N$ of volume $N/\rho$ can be placed in a cube $\Omega_N'$ of volume $N/\rho'$ when $\rho'<\rho$, an optimal position of the particles for $\Omega_N$ can be used in $\Omega_N'$, and we conclude that $\rho\mapsto e(\rho)$ is non-decreasing. Assume now that $V$ has an attractive part and that the problem $E(N)$ studied before in the whole space lead to a crystal. This crystal has an average density $\overline\rho=n/|Q|$ where $Q$ is the unit cell and $n$ is the number of particles in it ($n=1$ for a Bravais lattice). Then, if we impose a density $\rho<\overline{\rho}$ the domain will typically be larger than the natural size of the system and the particles will solve the exact problem $E(N)$, that is, we have $E_\Omega(N)=E(N)$. We conclude that $e(\rho)=e(\overline\rho)=e_\ii$ is constant for $0<\rho\leq \overline\rho$. The energy only starts to grow for $\rho>\overline\rho$.

The fact that one can consider a repulsive potential changes the physical meaning of the problem. In particular,
the relation with the sphere packing problem is less clear. In the case of the preceding section, it is natural to
consider that the particles are attracted to each other, and behave like hard spheres at short range, therefore
trying to maximize the density of the system. Doing so, they tend to maximize the number of their neighbors. Here,
particles can repel each other fiercely, and tend to maximize their mutual distance, while staying in the domain
$\Omega$. Experiments and numerical simulations indicate, however, that here again crystallization occurs.

In dimension one, Ventevogel and Nijboer have proved crystallization for any density $\rho>0$ in the case of
non-negative non-increasing convex potential~\cite{Ventevogel-78}. In \cite{VenNij-79-1,VenNij-79-2}, they prove
the same result for the potential $V(x) = \exp(-\alpha x^2)$ and $V(x) = (\beta + x^2)^{-1}$, for $\alpha,\beta>0$,
still in dimension one. This allows to generalize the result to any convex combination of these potential, such as
$V(x) = \int_0^{+\infty} e^{-\alpha x^2} d\mu(\alpha),$ for any non-negative measure $\mu$. Such potentials may be
non-convex. In addition, they give a necessary condition for crystallization, in any dimension: if crystallization
occurs for sufficiently small densities, and $V$ is continuous, then $\widehat V \geq 0$. As before, the situation
is much less clear in dimension $d\geq 2$.

\subsection{Optimal lattices and special functions}\label{sec:fonctions_speciales}

If crystallization is \emph{assumed}, it is possible to determine the most favorable periodic configurations by
comparing their energy per particle $e_\ii$ and $e(\rho)$, defined by~\eqref{eq:limite_thermo} and~\eqref{eq:limite_rho_fixee}. In
some cases, this question may be related to a problem in analytic number theory, involving special functions.

Indeed, if the particles lie on the vertices of a Bravais lattice $G$ (a discrete subgroup of $\R^d$ such
as~\eqref{eq:groupe_G}, with unit cell $Q$), the limit energy per particle reads:
\begin{equation}
\frac12\sum_{g\in G\setminus\{0\}}V(g).
\label{eq:energie_par_particule}
\end{equation}
Finding the optimal configuration amounts to minimize this expression with respect to $G$. There is no additional
constraint on $G$ for $e_\ii$. In contrast, one needs to fix the volume of the unit cell $Q$
of $G$ to $|Q|=1/\rho$ when the density is fixed, as for instance in the case of a repulsive potential.

\subsubsection*{Epstein zeta function}

With a Lennard-Jones type potential
\begin{equation}
  \label{eq:type-lj}
 V(r)=\frac 1 a \left(\frac{r_0}{r}\right)^{a}- \frac 1 b \left(\frac {r_0}{r}\right)^{b},
\end{equation}
where $a>b>d$, we get
\begin{equation}
\frac12\sum_{g\in G\setminus\{0\}}V_{\rm LJ}(g)=\frac{\zeta_d(S,a)}{a}-\frac{\zeta_d(S,b)}b
\label{eq:energy_epstein}
\end{equation}
where $S$ is a symmetric positive definite matrix of size $d$, which is related to the Gram matrix of the basis
$\left(v_j\right)_{1\leq j \leq d}$, and such that $r_0 S^{1/2}
\ZZ^d = G$. Here,
\begin{equation}
\zeta_d(S,s)=\frac12\sum_{z\in \Z^d\setminus\{0\}}\frac{1}{(z^TSz)^{s/2}}
\label{eq:zeta_Epstein}
\end{equation}
is the Epstein zeta function~\cite{Epstein-06}. Still assuming that we have crystallization on a Bravais lattice,
the minimal energy for $V_{\rm LJ}$ reads
\begin{equation}
e_{{\rm LJ},\zeta}=\min_{S=S^T>0}\left(\frac{\zeta_d(S,a)}{a}-\frac{\zeta_d(S,b)}b\right).
 \label{eq:valeur_min_V_LJ}
\end{equation}
Except in  dimension $d=1$, the solution to this problem is still unknown, even for the physically relevant cases $a=12$ and $b=6$.

If the density $\rho>0$ is fixed as discussed in Section~\ref{sec:densite_fixe}, it is possible to consider a
repulsive potential $V(r)=r^{-s}$ with $s>d$. Hence, we need to minimize the value of the zeta
function~\eqref{eq:zeta_Epstein}, with respect to  $G$ (that is, with respect to the matrix $S$)
\begin{equation}
e_\zeta(\rho,s)=\min_{\substack{S=S^T>0\\ \det(S)=\rho^{-2}}}\zeta_d(S,s).
 \label{eq:min_zeta}
\end{equation}
Here, $\det(S)$ is the volume of the unit cell of the lattice to the power $2$. Applying a dilation of the lattice,
one easily proves that
$$e_\zeta(\rho,s)=\rho^{s/d}\,e_\zeta(1,s)$$
and that it is sufficient to study the problem in which the unit cell has a volume equal to $1$. Without loss of
generality, we can thus assume that $\det(S)=1$. There is a link with the sphere packing since, in the limit
$s\to\ii$, the optimal lattice converges to a solution to the $d$-dimensional sphere packing
problem~\cite{Ryshkov-73}. It should be noted that $\zeta_d(S,s)$ is not bounded. If the smallest eigenvalue of $S$
reaches $0$, then $\zeta_d(S,s)$ tends to $+\ii$.

The function $s\mapsto\zeta_d(S,s)$ has an analytic continuation to the set $\C\setminus\{d\}$. This extension has a
simple pole at $s=d$, with a residue equal to $\pi^{d/2}\Gamma(d/2)^{-1}$
(if $\det(S)=1$), and satisfies the functional equation
\begin{equation}
 \zeta_d(S,s)=\pi^{s-d/2}\frac{\Gamma\left(\frac{d-s}{2}\right)}{\Gamma(\frac{s}2)}\zeta_d(S^{-1},d-s)
 \label{eq:formule_zeta}
\end{equation}
where $S^{-1}$  is the matrix associated with the lattice
$G^*=\left\{k\in\R^d \ :\ k\cdot g\in\Z,\ \forall g\in G \right\},$ called the
\emph{dual lattice of $G$}~\cite{BorGlaMPh-13,BorBorStr-14}. Thus, it is also possible to study the minimization
problem~\eqref{eq:min_zeta} even if $0<s<d$. As we will see below, this problem is of great importance from a
physical point of view, particularly if $d=3$ and $s=1$. Formula~\eqref{eq:formule_zeta} implies that if $S$ is
a solution to the minimization problem $e_\zeta(1,s)$, then $S^{-1}$ is a solution to $e_\zeta(1,d-s)$.

Going back to the case of the Lennard-Jones potential~\eqref{eq:valeur_min_V_LJ}, we see that, after dilating the
problem with fixed density $\rho>0$, it amounts to minimize the function
\begin{equation}
e_{{\rm LJ},\zeta}(\rho)=\rho^{a/d}\min_{\det(S)=1}\left(\frac{\zeta_d(S,a)}{a}-\rho^{\frac{b-a}{d}}\frac{\zeta_d(S,b)}b\right).
\label{eq:LJ_zeta}
\end{equation}
Since $b>a$ is assumed, for large $\rho$ the problem reduces to the
minimization of $\zeta_d(S,a)$.  For the minimization problem~\eqref{eq:valeur_min_V_LJ} with no density constraint, one finally
needs to consider $\overline\rho$ which minimizes the function $\rho\mapsto e_{{\rm LJ},\zeta}(\rho)$. In this case, the optimal lattice is unknown.

One can derive a representation of $\zeta_d$ as a series with exponentially decaying coefficients. The most widely
used method is due to Ewald~\cite{Ewald-21,BonMar-77,GiuVig-05} and relies on the integral representation
\begin{equation}
\frac{1}{r^s}=\frac{1}{2\Gamma(s/2)}\int_0^\ii e^{-\tau r^2}\tau^{s/2-1}\,d\tau.
\label{eq:representation_integrale}
\end{equation}
For $s>d$, we have, if $\det(S)=1$,
\begin{equation}
\zeta_d(S,s)=\frac{\pi^{s/2}}{\Gamma(s/2)}\left\{\frac{1}{s-d}-\frac{1}{s}+\frac12\int_1^\ii \Big(\big(\theta_d(S,\tau)-1\big)\tau^{s/2-1}+\big(\theta_d(S^{-1},\tau)-1\big)\tau^{\frac{d-s}{2}-1}\Big)\,d\tau\right\},
\label{eq:formule_zeta_theta}
\end{equation}
where
\begin{equation}
\theta_d(S,\alpha)=\sum_{z\in \Z^d} e^{-\pi \alpha z^TSz}.
\label{eq:theta_Jacobi}
\end{equation}
is the \emph{Jacobi theta} function. Here, we have used Poisson's summation formula
\begin{equation}
  \label{eq:theta_invariance}
  \theta_d(S,\alpha) = \frac 1 {\alpha^{d/2}} \theta_d\left(S^{-1}, \frac
    1 \alpha\right).
\end{equation}

Formula~\eqref{eq:formule_zeta_theta} is also meaningful for $0<s<d$ and can be used to prove that $\zeta_d$ has
an analytic extension to $\C\setminus\{d\}$ ($\Gamma$ has a pole at the origin which compensates for the divergent term
$1/s$), as we already mentioned. Formula~\eqref{eq:formule_zeta_theta} is widely used by physicists. It allows to compute numerically the values of
$\zeta_d(S,s)$ very accurately, allowing to formulate conjectures on what should be proved.

We are now going to describe what is expected for the minimization of the Epstein zeta function.

\subsubsection*{Results for $\zeta$ and $\theta$ in dimension 2}
In dimension $d=2$, it has been proved by Rankin~\cite{Rankin-53}, Cassels~\cite{Cassels-59},
Ennola~\cite{Ennola-64} and Diananda~\cite{Diananda-64}, that the hexagonal lattice is the unique minimizer of the zeta
function, for any $s>0$, when the density is fixed. In other words, we have
\begin{equation}
\zeta_2(S,s)-\zeta_2(S_{\rm hex},s)\geq0
\label{eq:zeta_2D}
\end{equation}
for all $s>0$ and all $S$ such that  $\det(S)=1$, where
$$S_{\rm hex}=\frac{2}{\sqrt{3}}\begin{pmatrix}
1 & 1/2\\
1/2 & 1
    \end{pmatrix}
$$
corresponds to the hexagonal lattice. In addition, the inequality~(\ref{eq:zeta_2D}) is strict if $S\neq S_{\rm
  hex}$, up to the invariances of the problem (rotation and change of basis of the lattice). Another proof is given
in~\cite{NonVor-98}. Inequality~\eqref{eq:zeta_2D} is still valid for $s=2$ where both functions have a simple pole
with equal residue. When $s\to0$, we have a divergence which needs to be dealt with, but the result is still true~\cite{SanSer-12}. We have made some numerical calculations that confirm these results, see Figure~\ref{fig:carre_vs_hexagonal} below.

A famous result due to Montgomery~\cite{Montgomery-88} deals with the case of a Gaussian repulsive interaction in
dimension $d=2$. In this case, the problem reduces to the study of the Jacobi theta
function~\eqref{eq:theta_Jacobi}. As for the zeta function, Montgomery proves that $\theta_2(S,\alpha)\geq \theta_2(S_{\rm hex},\alpha)$
for all $\alpha>0$ and all $S$ such that $\det(S)=1$ (cf. Figure~\ref{fig:carre_vs_hexagonal_theta}). If the
potential $V$ is a positive linear combination of Gaussians, Montgomery's result implies that the optimal lattice
is the hexagonal one, for a fixed unit cell volume. For instance, using the integral
formula~\eqref{eq:formule_zeta_theta}, one recovers the previously mentioned result on zeta functions.

Subtracting two Epstein zeta functions gives a function that can be expressed as an integral of the function
$\theta_2(S,\alpha)$ multiplied by a weight. This weight is non-negative when $\rho$ is large enough. Using this
argument, Bétermin and Zhang~\cite{BetZha-14} have proved that, at high density, the optimum of~\eqref{eq:LJ_zeta} is reached by the hexagonal lattice in
2D, for the Lennard-Jones potential $V_{\rm LJ}$. Imposing that $\rho$ is large means that particles are close to
each other. Therefore, their interaction is dominated by the repulsive part $r^{-12}$ of $V_{\rm LJ}$. The energy
is close to $\zeta_2(S,12)$, which, as a function of $S$, reaches its minimum for the hexagonal lattice only. On
the contrary, they prove that, when $\rho\to 0$, the hexagonal lattice cannot be the global minimizer. For
instance, if
$$\rho^{3}<\frac{\zeta_2(I_2,6)-\zeta_2(S_{\rm hex},6)}{\zeta_2(I_2,12)-\zeta_2(S_{\rm hex},12)},$$
the square lattice ($S_{\rm car}=I_2$) has an energy which is smaller than that of the hexagonal lattice. If no
symmetry breaking occurs, then the square lattice becomes the minimizer. Recall, however, that the true $e(\rho)$ is expected to be constant when $\rho\leq\overline\rho$ (the density of the optimal lattice at which~\eqref{eq:LJ_zeta} is minimal). It will not coincide with $e_{{\rm LJ},\zeta}(\rho)$ in~\eqref{eq:LJ_zeta} at small $\rho$. For a more recent work in the same spirit, see~\cite{Betermin-15}.

\begin{figure}[ht]
\centering
\includegraphics[width=7.5cm]{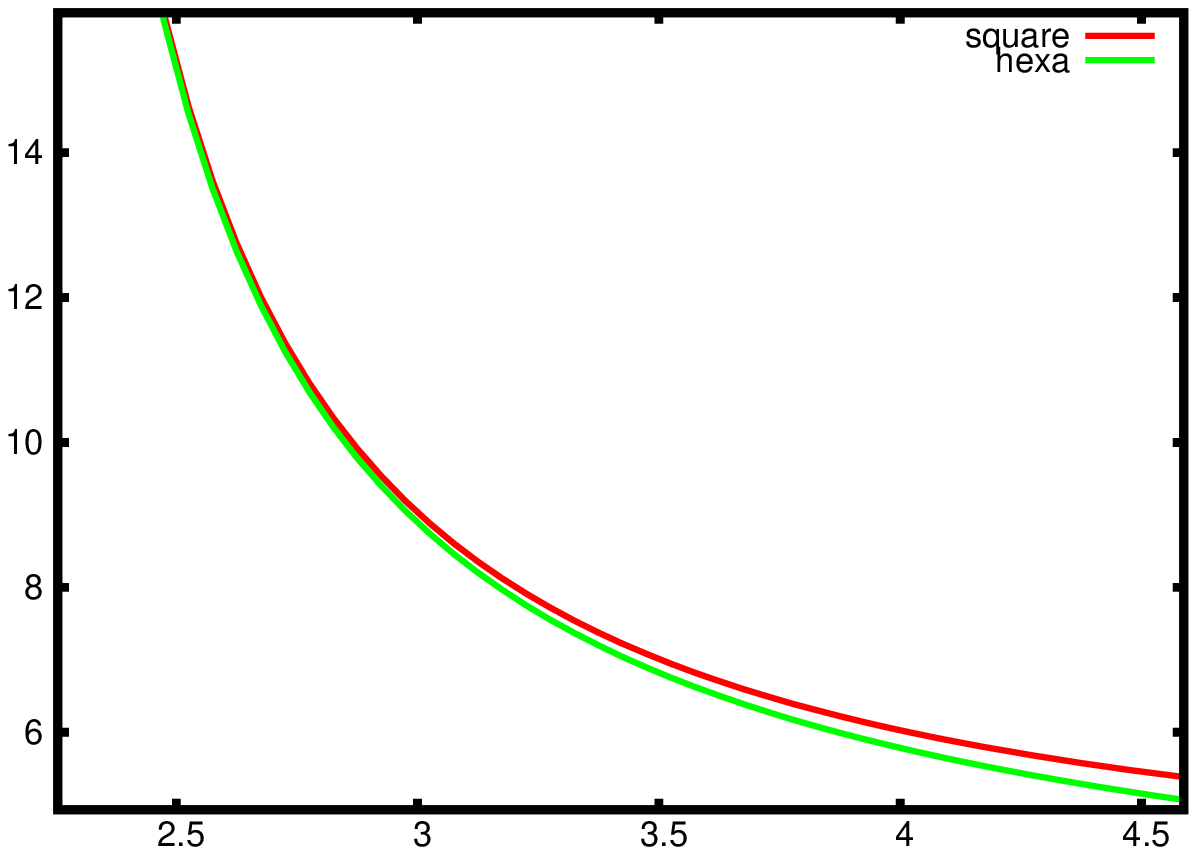}\includegraphics[width=7.5cm]{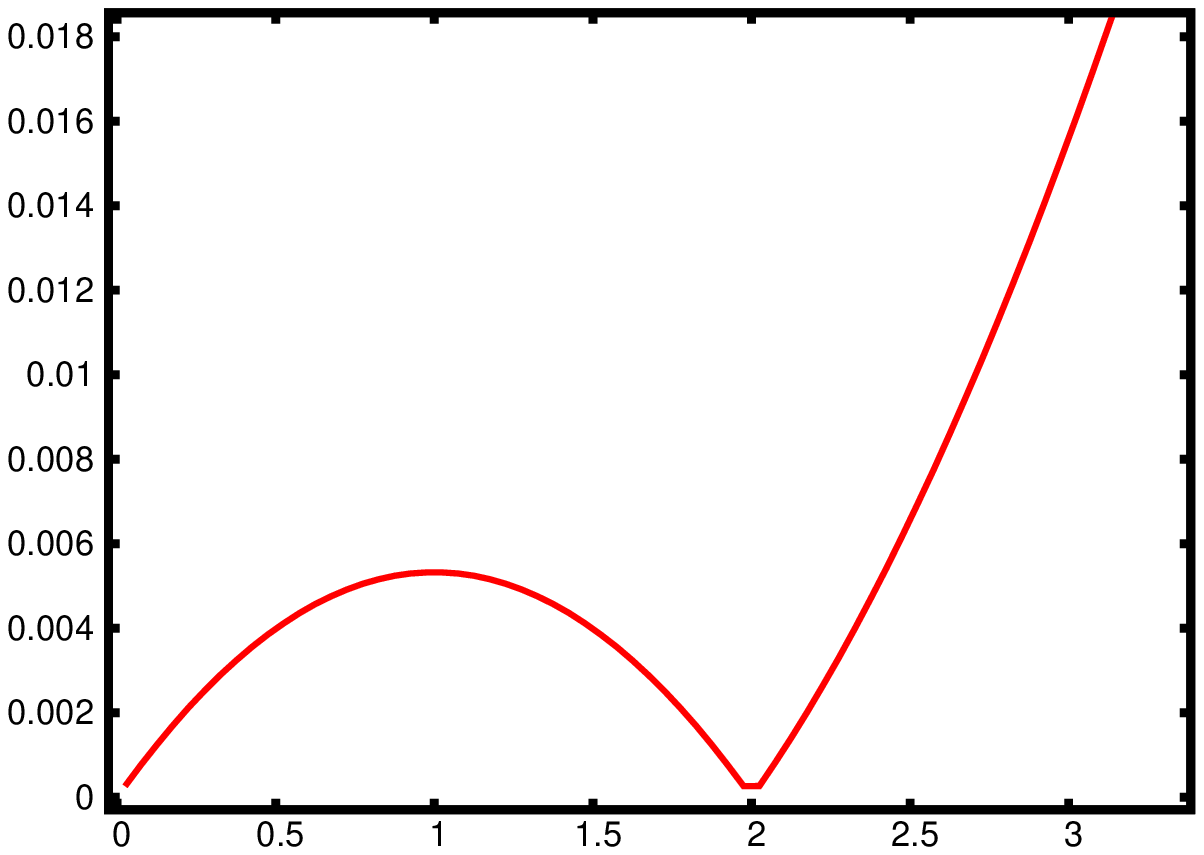}
\caption{\small Left: $\zeta_2(S,s)$ as a function of $s$ for the square lattice ($S=I_2$) and the hexagonal one
  ($S=S_{\rm hex}$). Right: the relative difference $\big(\zeta_2(I_2,s)-\zeta_2(S_{\rm
    hex},s)\big)/\left|\zeta_2(S_{\rm hex},s)\right|$. It shows that the hexagonal lattice energy is lower than
  that of the square lattice for all $s>0$, as it is proved in~\cite{Rankin-53,Cassels-59,Ennola-64,Diananda-64}.
\label{fig:carre_vs_hexagonal}}

\medskip

\includegraphics[angle=270,width=7.5cm]{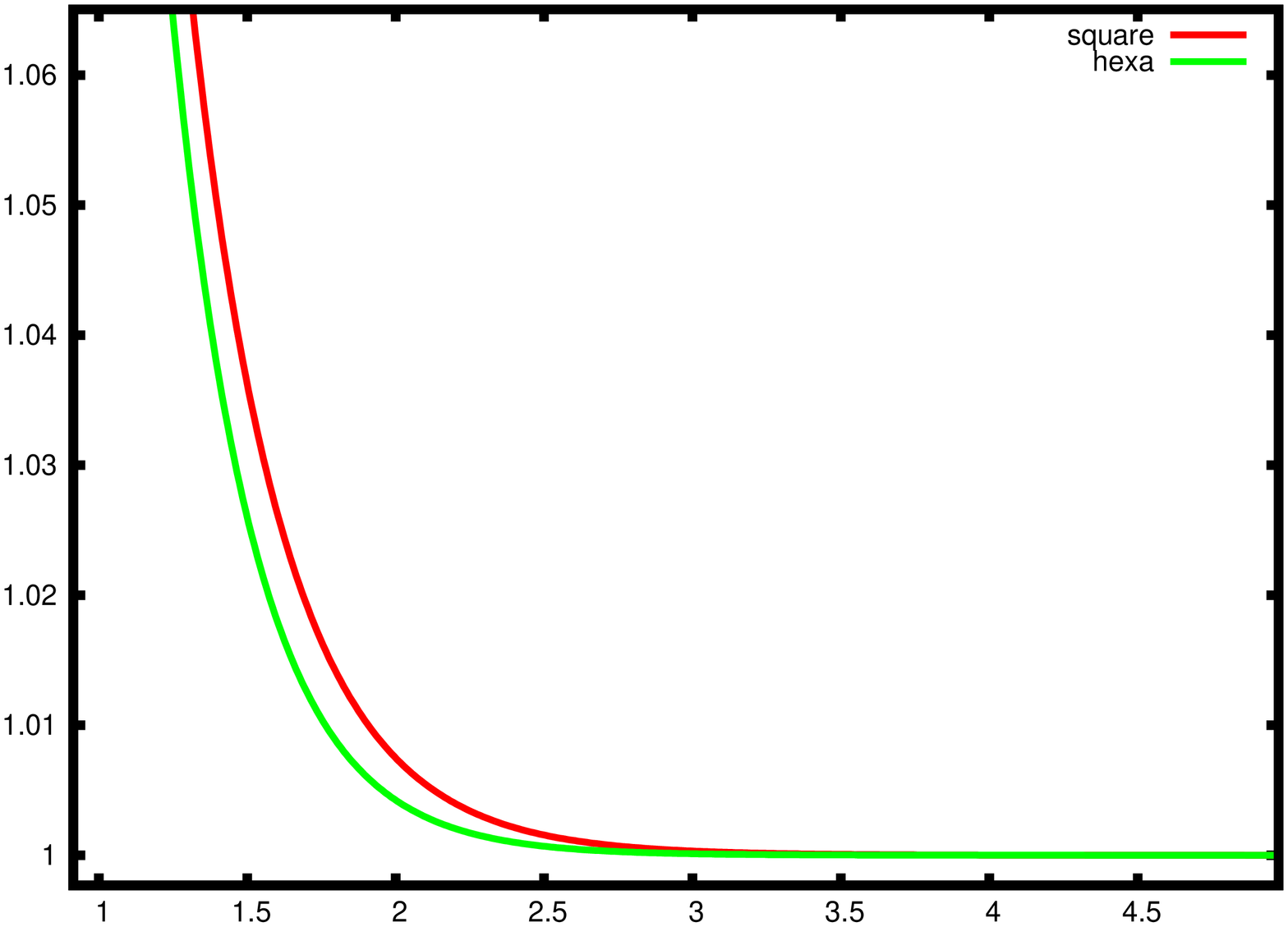}\includegraphics[angle=270,width=7.5cm]{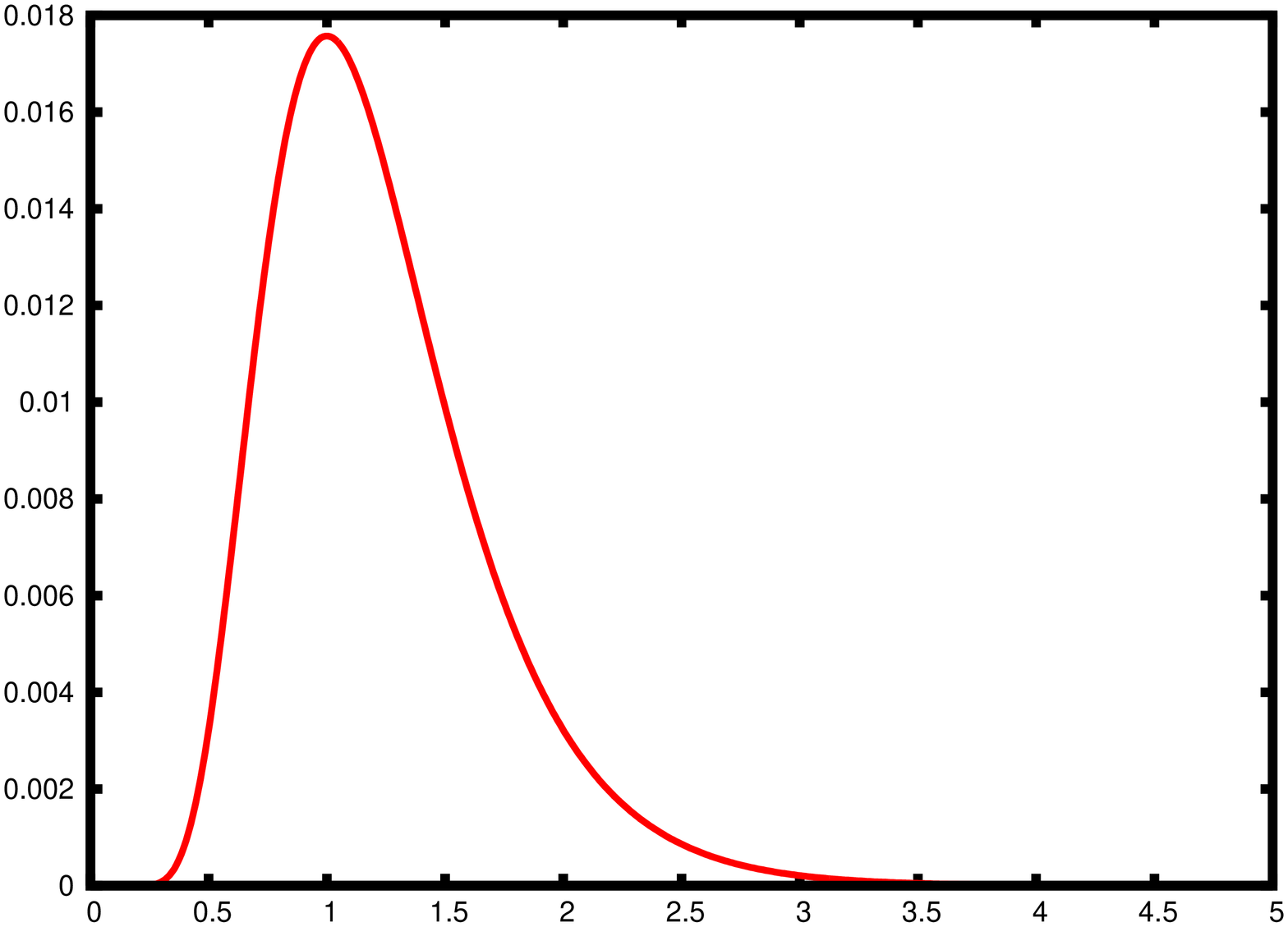}
\caption{\small Left: $\theta_2(\alpha)$ as a function of $\alpha$ for the square lattice ($S=I_2$) and the hexagonal one
  ($S=S_{\rm hex}$). Right: the relative difference $\big(\theta_2(I_2,\alpha)-\theta_2(S_{\rm
    hex},\alpha)\big)/\theta_2(S_{\rm hex},\alpha)$. It shows that the hexagonal lattice energy is lower than
  that of the square lattice for all $s>0$, as it is proved in~\cite{Montgomery-88}.
\label{fig:carre_vs_hexagonal_theta}}
\end{figure}

\subsubsection*{Results and conjectures for $\zeta$ and $\theta$ in dimension $d\geq3$}

\begin{figure}[h]
\centering
\includegraphics[width=7.5cm]{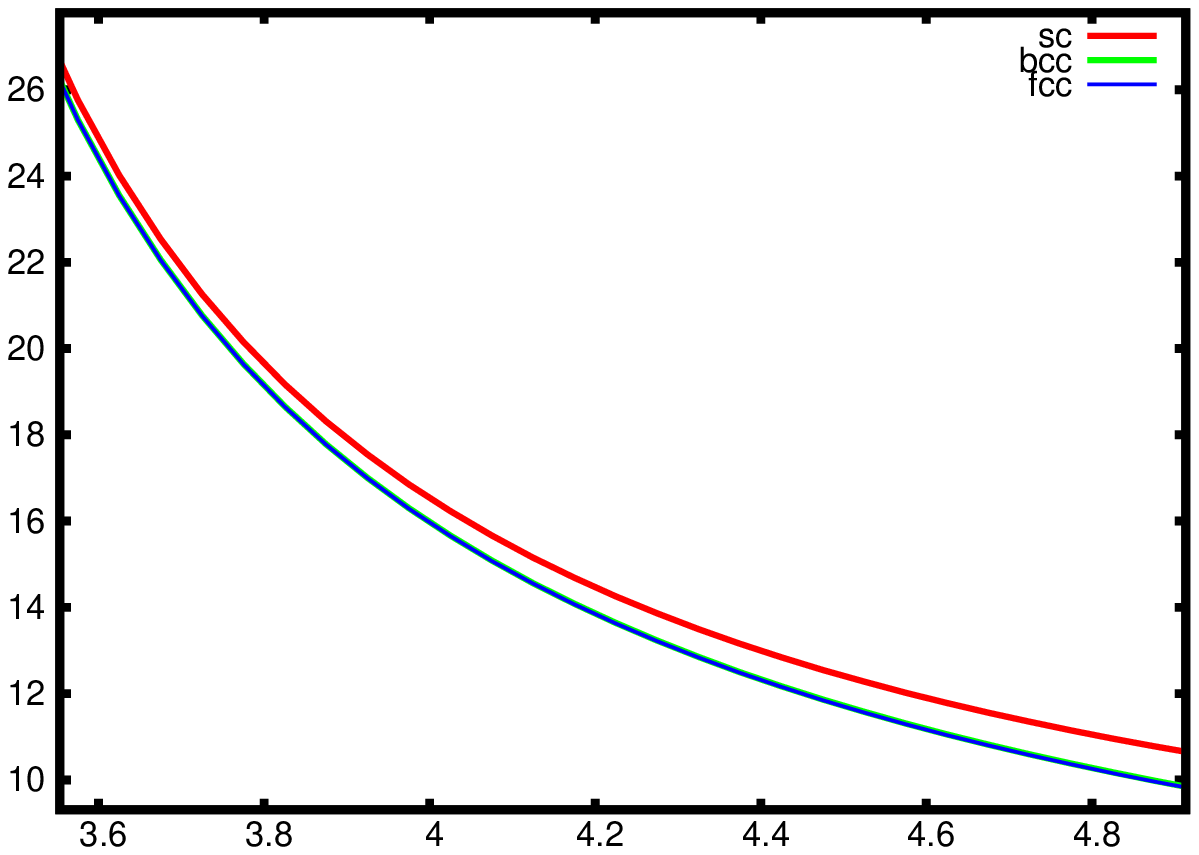}\includegraphics[width=7.5cm]{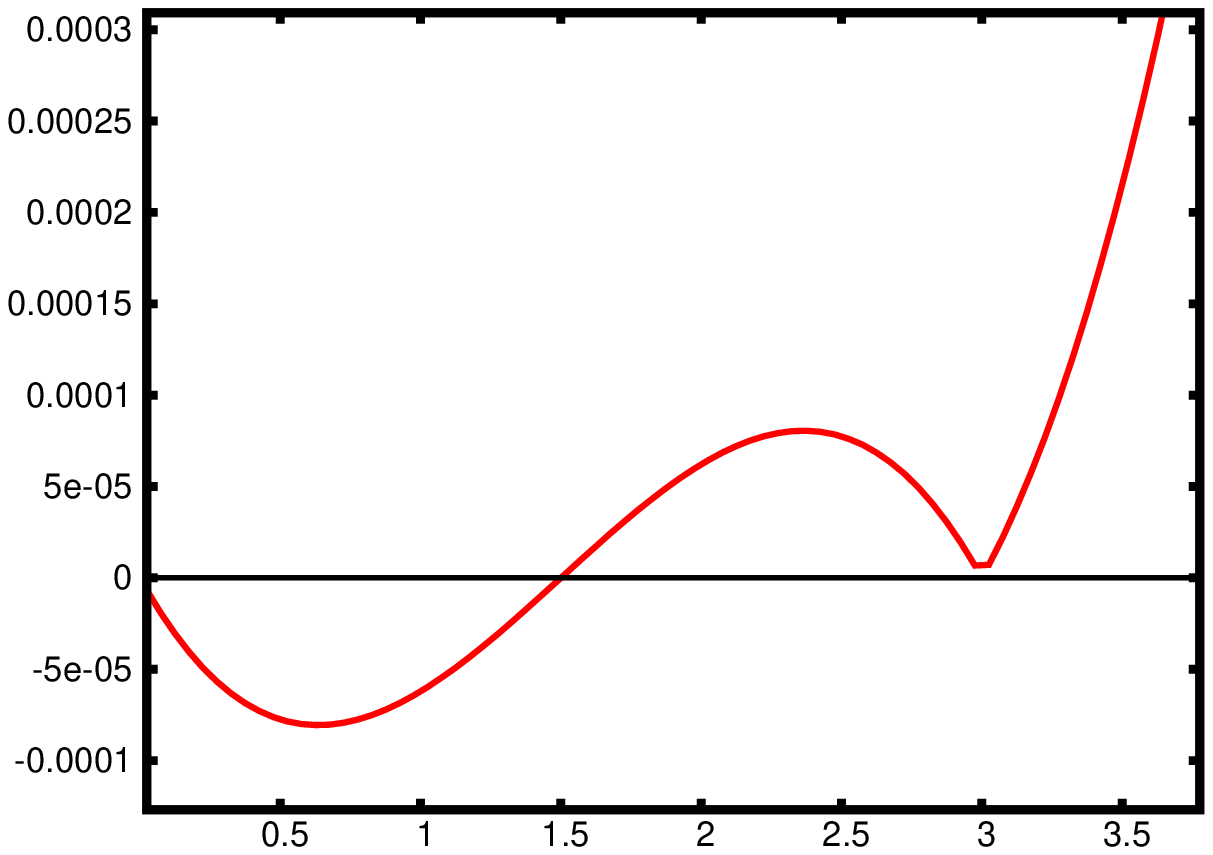}
\caption{\small Left: $\zeta_3(S,s)$ as a function of $s$ for different lattices ; FCC and BCC have energies which
  are very close to each other. Right: the relative difference $\big(\zeta_3({\rm BCC},s)-\zeta_3({\rm
    FCC},s)\big)/\left|\zeta_3({\rm FCC},s)\right|$. It indicates that BCC should be the minimizer for $0<s<3/2$, while FCC
  should be for $s>3/2$. Proving this is still an open problem. The relative difference is of order $10^{-4}$.
\label{fig:BCC_vs_FCC}}

\medskip

\includegraphics[angle=270,width=7.5cm]{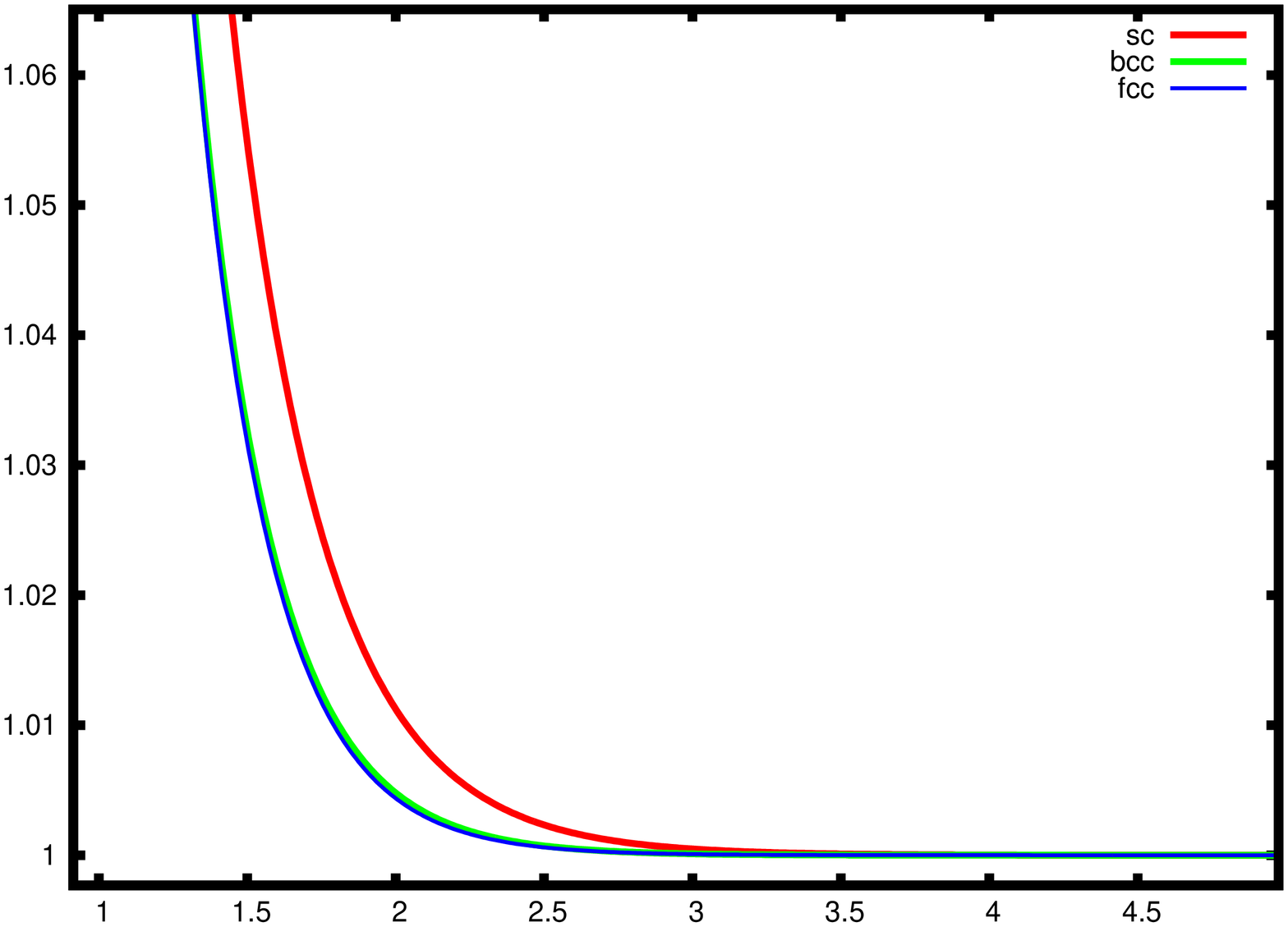}\includegraphics[angle=270,width=7.5cm]{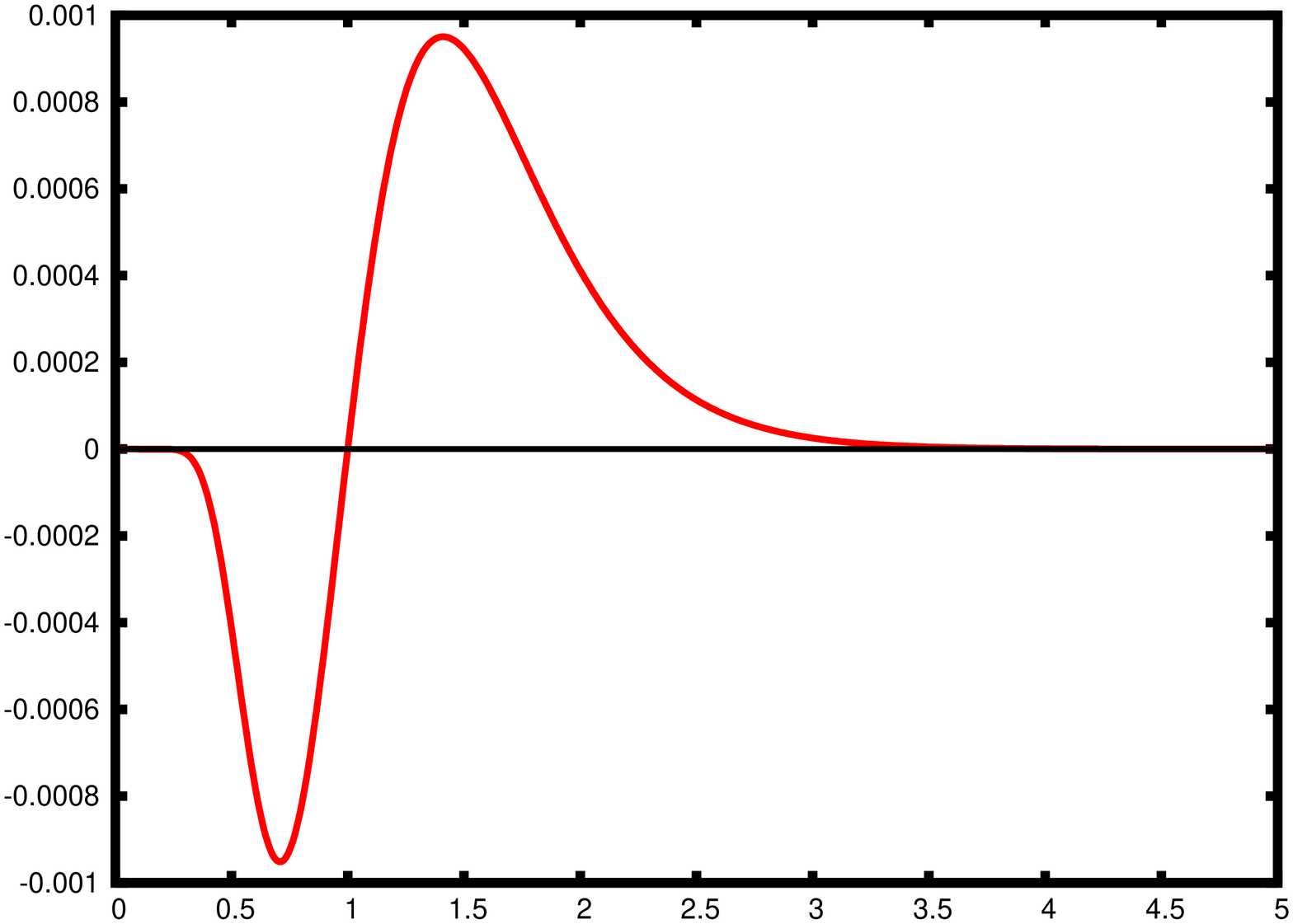}

\caption{\small Left: $\theta_3(\alpha)$ as a function of $\alpha$ for several lattices ; FCC and BCC have energies which
  are very close to each other. Right: the relative difference $\big(\theta_3({\rm BCC},\alpha)-\theta_3({\rm
    FCC},\alpha)\big)/\theta_3({\rm FCC},\alpha)$. It indicates that FCC is the minimizer for $\alpha>1$, while BCC
  is for $\alpha<1$. Proving this is an open problem, which does not seem to imply the above result on the function $\zeta_3$.
  \label{fig:theta_3d}}
\end{figure}

In dimension $d\geq 3$, some authors have studied the critical points and the (local or
global) minima of the Jacobi theta function and the Epstein zeta function. In a famous article~\cite{SarStr-06}, Sarnak and
Strömbergsson determined special local minima in dimensions 4, 8 and 24 (see
also~\cite{Coulangeon-06,CohKum-07,CouSch-12}). In dimension 3, Ennola has proved that the face centered cubic
(FCC) lattice is a non-degenerate local minimum of $\zeta_3(S,s)$ for all
$s>0$~\cite{Ennola-64b}. Formula~\eqref{eq:formule_zeta} implies that its dual, the BCC lattice, is also a
non-degenerate local minimum for $0<s<3$. In addition, based on the sphere packing problem obtained in the limit $s\to\ii$, it has been shown in~\cite{Ryshkov-73} that FCC is the unique global minimizer for $s$ large enough.
As opposed to what Ennola conjectured in~\cite{Ennola-64b}, FCC
cannot be the unique minimizer for all $s>0$. Indeed, formula~\eqref{eq:formule_zeta} would imply that its dual, BCC,
is a minimizer for some values of $s$. Hence, a more likely conjecture would be that FCC is the unique minimizer
for $s>3/2$, whereas BCC is for $0<s<3/2$ \cite[section 5]{SarStr-06}.\footnote{Note that BCC cannot be a non-degenerate local minimum for all $s>0$~\cite{Fields-80,Gruber-12}, hence it has to be a degenerate critical point of $\zeta_d$ for some $s\geq3$.} If it is assumed that the minimizer has a
high-symmetry group, and if we only compare the energies of SC, FCC and BCC, this conjecture is corroborated by
numerical computations presented in Figure~\ref{fig:BCC_vs_FCC}. It is a very important conjecture: its proof would
be an important advance both in analytic number theory and in solid-state physics. One of the difficulties in the
proof is that the values of the zeta function for BCC and FCC are very close to each other. This implies that a
quantitative argument needs to be very precise.

Similar questions may be asked about theta function~\eqref{eq:theta_Jacobi}, but it seems that the corresponding
literature is far less important. In 3D, the conjecture is that FCC is the unique minimizer for any $\alpha>1$,
whereas BCC is for $\alpha<1$~\cite[section 5]{SarStr-06}. Here again, this conjecture is confirmed by numerical
simulations presented in Figure~\ref{fig:theta_3d}. Note that, contrary to dimension 2, the conjecture for the theta
function does not seem to imply it for the zeta function: formula~\eqref{eq:representation_integrale} always involves both the
lattice and its dual for different values of $\alpha$.

Most works consider only mono-atomic lattices. This excludes the HCP (Hexagonal Close Packed) lattice in dimension
3, since it is not a Bravais lattice. We refer to~\cite{LimTeo-08} for an explicit link between zeta functions and
quantum field theory, to~\cite{Sobolev-61} for the link with optimal quadrature point repartition, and
to \cite{OsgPhiSar-88,Sarnak-90,SarStr-06} for the link with the optimization of the determinant of the Laplace
operator: $\det(-\Delta)=e^{-\zeta'_d(S,0)}$.

As a conclusion, determining the optimal periodic lattice can, for some simple potentials, be related to the study
of special functions. the conjecture is that the minimizer can be either the FCC lattice, or the BCC one. This is still
an open problem (in most cases), even though research is very active on this subject.

\subsection{Coulomb potential and Wigner crystallization}
\label{sec:Wigner}

The Coulomb potential (the real interaction between charged particles) is not covered by any of the previous results. In dimension $d$, the Coulomb potential $V_{d,\rm Coul}$ is by definition the Green function of the Laplace operator, that is, the
solution to
$$-\Delta V_{d,\rm Coul}=|S^{d-1}|\delta_0,$$
in the sense of distribution, where $|S^{d-1}|$ is the volume of the sphere in dimension $d$, $S^{d-1}=\{x\in\R^d\
:\ |x|=1\}$. We thus have
$$V_{1,\rm Coul}(x)=-|x|,\qquad V_{2,\rm Coul}(x)=-\log|x|,\qquad V_{d,\rm Coul}(x)=\frac{1}{|x|^{d-2}}\quad \text{for $d\geq3$}.$$
The function $V_{d,\rm Coul}$ is never integrable at infinity and it is not stable in dimension $d\leq 2$. In dimension $d\geq3$, $V_{d,\rm Coul}$ is non-negative, hence no macroscopic object can be formed. If we perform the thermodynamic limit as in Section~\ref{sec:densite_fixe}, then $E(\Omega_N,N)$ behaves as $N^{1+2/d}$.

Physically, a macroscopic system of charged particles is never seen in vacuum. Two alternatives are usually
considered. The first is to put them in a trap, that is, to add an external potential $\sum_{j=1}^NV_{\rm ext}(x_j)$ to the energy, with $V_{\rm ext}(x)\to+\ii$ when $|x|\to\ii$. This is now done in the laboratory~\cite{ChuLin-94,HayTac-94,Thoetal-94,Thompson-15}, although a large number of particles is still difficult to reach. We
discuss this possibility in Section~\ref{sec:riesz} below. Another realistic macroscopic system of electrons is
when they are placed in an external potential describing a background of opposite charge, which compensates the
charges of the electrons and allows for an equilibrium to form. In a metal, the background is composed of the
positively-charged nuclei (and possibly another set of electrons that do not move). These other particles can be
fixed or optimized (see Section~\ref{sec:several_types}). In~\cite{Wigner-34}, Wigner considered the simpler
situation of a \emph{uniform background} of density $\rho>0$, and this is the object of this section. This is the
so-called \emph{Jellium} model, in which the background is a kind of ``jelly'' slowing down the movements of the
particles.  As we will explain in Section~\ref{sec:riesz} below, trapped systems can be shown to locally behave like Jellium hence, in spite of its apparent simplicity, the Jellium problem is of high physical relevance.

For a general potential $V$, the Wigner minimization problem reads:
\begin{multline}
E_{\Omega,\rho}(N)=\!\!\inf_{x_1,\dots,x_N\in \Omega} \Bigg\{\sum_{1\leq i < j\leq N}
V\left(|x_i- x_j|\right) -\rho\sum_{i=1}^N \int_\Omega V(|x_i-y|)\,dy\\+\frac{\rho^2}{2}\int_\Omega\int_\Omega V(|x-y|)\,dx\,dy \Bigg\},
\label{eq:min_Coulomb}
\end{multline}
with the same limit as before
\begin{equation}
e_{\rm Jell}(\rho)=\lim_{\substack{N\to\ii\\ |\Omega_N|\to\ii\\ N/|\Omega_N|\to\rho}}\frac{E_{\Omega_N,\rho}(N)}{N}.
 \label{eq:limit_Coulomb}
\end{equation}
The second term of the energy in~\eqref{eq:min_Coulomb} accounts for the interaction of our $N$ particles with the
homogeneous background. This is a new term compared to the preceding cases. The last term is the energy of the
background, which is constant with respect to the positions of the particles. We keep it in order to have a finite
limit~\eqref{eq:limit_Coulomb} (in the case we deal with here, $\int_\Omega\int_\Omega V(|x-y|)\,dx\,dy$ grows
like $N^{1+2/d}$). In the limit~\eqref{eq:limit_Coulomb} one imposes that
$N/|\Omega_N|\to\rho$, which means that the particle density is equal to that of the background. This allows to
reach an electrostatic equilibrium between particle repulsion and the attraction of the background. One could in
principle minimize over the domain $\Omega$ while fixing $|\Omega|=N\rho$, and then use this domain $\Omega_N$, but
it is often assumed that $\Omega_N$ has a shape which is fixed (a cube or a ball for instance), and is then
dilated. The limit~\eqref{eq:limit_Coulomb} should not depend on the chosen sequence $\Omega_N$.

For a stable integrable potential $V$, adding the background should only affect the points close to the boundary of $\Omega_N$, and we expect that in the thermodynamic limit $e_{\rm Jell}(\rho)=e(\rho)-\rho/2\int_{\R^d}V$. The situation is different for potentials that decay slowly. As mentioned above, Wigner was originally interested in the electrostatic interaction between electrons, that is, the Coulomb potential $V=V_{d,\rm Coul}$. Nevertheless, the problem~\eqref{eq:min_Coulomb} makes sense as soon as $V$ behaves like $|x|^{-s}$ with $s\geq d-2$ at infinity. \emph{Screening} is the main effect that will make this possible. The idea is that each particle will feel an effective potential that decays faster than $|x|^{-s}$, due to the cancellations induced by the two additional terms. The potential is expected to decay like $|x|^{-s-2}$ or even $|x|^{-s-3}$ if the configuration of the particles has sufficiently many symmetries.

In~\cite{Wigner-34}, Wigner has conjectured crystallization for $V=V_{d,\rm Coul}$, at least if $1\leq d\leq3$ and $\rho$ is small enough. He
also suggested that in 3D, the electrons form a body centered cubic lattice (BCC). In 2D, the particles are
expected to form a hexagonal lattice. The same is expected for $V(|x|)=|x|^{-s}$ with $d-2\leq s<d$. Numerical simulations and formal computations corroborate Wigner's conjecture. However, a rigorous proof is still missing in dimensions $d= 2,3$.

Dimension $d=1$ is simpler and has been solved by Kunz in 1974 for small densities~\cite{Kunz-74}. This result has
been generalized to any density by Aizenman and Martin~\cite{AizMar-80}. At temperature $T=0$, the particles form a
lattice of step $1/\rho$, as in the preceding sections. If $T>0$, it has also been proved in~\cite{AizMar-80} that the
particle density is periodic of period $1/\rho$. A different proof, which applies to the quantum case, has been
proposed by Brascamp and Lieb in~\cite{BraLie-75}. It is an application of their study of the optimality of Gaussians in some
functional inequalities.

As in Section~\ref{sec:fonctions_speciales}, if crystallization is assumed (on a Bravais lattice $G$), it is
possible to compute the corresponding energy per unit volume. In the present case, we have
\begin{equation}
e_{{\rm Jell},G}(\rho)=\frac12\sum_{g\in G\setminus\{0\}}W(g)-\int_Q V(x)\,dx+\frac{\rho}2\int_Q\int_Q V(x-y)\,dx\,dy
\label{eq:valeur_energie_Coul}
\end{equation}
where $W$ is the twice-screened potential
$$W(x)=V(x)-2\frac{1}{|Q|}\int_{Q}V(x-y)\,dy+\frac{1}{|Q|^2}\int_{Q}\int_{Q}V(x+y-z)\,dy\,dz$$
with $Q$ the unit cell of the lattice $G$, satisfying $|Q|=1/\rho$.
As expected, the two terms have the same effect as a Taylor expansion of $V$ at infinity, which increases its decay and can make the series in~\eqref{eq:valeur_energie_Coul} convergent (depending on the symmetry properties of the unit cell $Q$). If $V(x)=|x|^{-s}$ at infinity with $s>d-2$, then $W(x)\sim |x|^{-s-2}$ and no assumption on $Q$ is necessary. For $s=d-2$ (Coulomb potential), it is sufficient to choose for $Q$ a set which is symmetric with respect to the origin $0$, which is always possible. Doing so, $W$ behaves like $|x|^{-d-1}$ at
infinity. If $Q$ has sufficient symmetry properties, it is possible to express the energy with the simple screened potential
$$\tilde W(x)=V(x)-\frac{1}{|Q|}\int_{Q}V(x-y)\,dy.$$
We refer to ~\cite[App.~B]{LewLie-15} for the details. It is always useful to choose for $Q$ the Wigner-Seitz cell,
which has the same symmetries as the lattice $G$~\cite{AshcroftMermin}.

When the above series converges and $V(x)=|x|^{-s}$, one can prove that the energy~\eqref{eq:valeur_energie_Coul} is \emph{equal to the
  analytic extension of the first term in~\eqref{eq:min_Coulomb}}, that is,
\begin{equation}
\boxed{e_{{\rm Jell},G}(\rho)=\zeta_d(S,s),}
\end{equation}
for $d>2$ and $d-2\leq s<d$. The proof is based on the same arguments as~\cite[App.~B]{LewLie-15}, and on results
from~\cite{BorBorShaZuc-88,BorBorSha-89,BorBorStr-14}. The point $s=d-2$ which is the one interesting for applications is always on the left of the pole $s=d$, where the Zeta function is defined through analytic continuation, showing the importance of studying it in this region.
For $d=3$, the numerical simulations presented in this article indicate that the body centered cubic lattice (BCC)
is the unique global minimizer, as conjectured by Wigner. Proving this fact is still an open problem. In dimension
$d=2$, the energy has a logarithmic singularity which needs to be removed, but the problem is
similar. In~\cite{SanSer-12}, Sandier and Serfaty used Montgomery's result to prove that the optimal lattice in 2D
must be the hexagonal lattice, in the limit $s\to 0$.

Physicists usually rely on the integral representation~\eqref{eq:representation_integrale} to compute the value of
the zeta function and compare different lattices. As an example, the energies are approximately equal to
$$-\rho^{1/3}\begin{cases}
1.41865...&\text{for the simple cubic lattice (SC),}\\
1.44415...&\text{for the face centered cubic lattice (FCC),}\\
1.44423...&\text{for the body centered cubic lattice (BCC),}
\end{cases}$$
in dimension $d=3$~\cite{BorGlaMPh-13,GiuVig-05}. Using an argument due to Onsager~\cite{Onsager-39}, Lieb and
Narnhofer managed to prove in~\cite{LieNar-75} that the true energy defined by~\eqref{eq:limit_Coulomb} satisfies
$e_{\rm Jell}(\rho)\geq -\rho^{1/3}1.4508...$ for any $\rho>0$. This value is very close to the expected
one. However, the proof of Wigner crystallization is still an open problem, in dimension $d\geq2$.

Let us point out that Wigner's model has been recently studied and reformulated
in~\cite{SanSer-14,RouSer-14,PetSer-14}. In these articles, the energy $e_{\rm Jell}(\rho)$ is called
\emph{renormalized energy} and is defined directly on sets of infinitely many points (which need not be on a periodic lattice), without using
the thermodynamic limit $N\to\ii$.

\subsection{Occurrence of the crystallization problem in other situations}
\label{sec:pb_relies}

In this section, we present a few questions that reduce to the crystallization problem stated above, or
to Wigner's problem. This shows that these questions are rather universal.

\subsubsection{Confined systems in the mean-field limit}
\label{sec:riesz}

The crystallization problem appears in some dense trapped systems that are now realized in the laboratory. Here, a change of scale is needed to recover a
problem set in the whole space. The prototypical situation is to minimize the energy
\begin{equation}
  \label{eq:riesz}
E_{V_{\rm ext}}(N)=\min_{x_i}\left\{\frac 1 N  \sum_{1\leq i<j\leq N} V(|x_i -x_j|) +  \sum_{i=1}^N V_{\rm ext}(x_i)\right\},
\end{equation}
where $V_{\rm ext}$ is a confining potential, which tends to $+\infty$ at infinity. The coefficient $1/N$
multiplying the interaction allows for both terms to be of the same order of magnitude in the limit
$N\to+\infty$. This is called the \emph{mean field} regime. An example of interaction $V$ is given by $V(x) =
|x|^{-s}$ with $0<s<d$, or $V(x) = -\log|x|$. Most commonly used confining potentials are the harmonic potential
$V_{\rm ext}(x) = |x|^2$, and the potential
\begin{displaymath}
  V_{\rm ext} (x) =
  \begin{cases}
    0 &\text{if } x\in \cM, \\
    +\infty & \text{otherwise,}
  \end{cases}
\end{displaymath}
which amounts to impose that all the particles stay in the bounded set $\cM\subset \R^d$. The set $\cM$ can be a bounded
domain like a ball, or a zero-measure set such as a sub-manifold of $\R^d$, of dimension strictly smaller than $d$.

\begin{figure}[t]
\centering
\includegraphics[width=10cm]{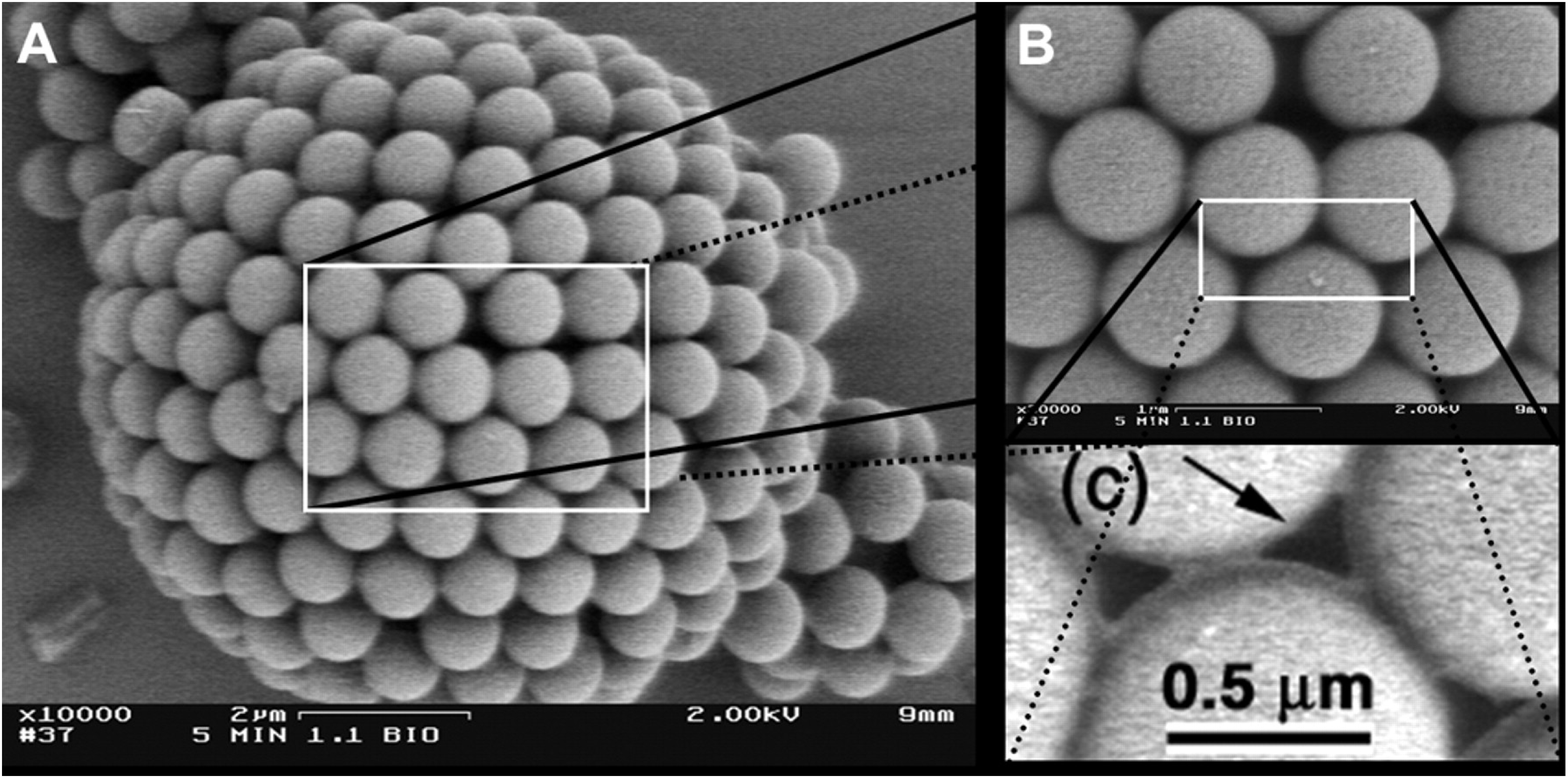}
\caption{\small A colloidosome is a spherical shape made of colloids (polystyren molecules here), which is
  described by Thomson's model. This model consists in minimizing the interaction $|x|^{-s}$ for particles on the
  sphere $S^2\subset\R^3$. Scanning microscope picture, from~\cite{Dinetal-02}. {\scriptsize\copyright\ AAAS.}\label{fig:colloidosomes}}
\end{figure}

If $d=3$ and $\cM = S^2$ is the unit sphere, and if $V(x)=  |x|^{-1}$, this is called \emph{Thomson's
  problem}~\cite{Thomson-04}. Finding the optimal positions of the particles on the sphere, even for a fixed value
of $N$, is a famous problem which has been solved only for some values of $N$. Many numerical studies of the
problem have been made, giving some insight on what the optimal configurations should look like. This problem is
related to one of the eighteen open problems mentioned by Smale in 1998~\cite{Smale-98,Beltran-13,Beltran-13b}. It naturally occurs in many different
situations: it is related to the construction of a set of points which discretizes the sphere as uniformly as possible (the so-called elliptic Fekete
points~\cite{SafTot-97}); in biology, this problem can explain the form of some viruses, and the repartition of
pores on pollen grains; it is also studied in link with ``colloidosomes''~\cite{Dinetal-02}
(Figure~\ref{fig:colloidosomes}). If $V(x)=|x|^{-s}$, with $d>s$, and if $\cM$ is a sub-manifold without boundary,
of dimension $d-1$, the problem is usually called \emph{Riesz problem}. We refer to~\cite{HarSaf-04} for a general
presentation of the problem and numerical simulations. Coulomb crystals in traps are now produced in the laboratory, since the 90s~\cite{ChuLin-94,HayTac-94,Thoetal-94,Thompson-15}.

For the model~\eqref{eq:riesz}, a second-order expansion is needed to find the crystallization problem. Indeed, the
leading order is, under appropriate assumptions, given by the mean-field theory
\begin{equation}
\lim_{N\to\ii}\frac{E_{V_{\rm ext}}(N)}{N}=\inf_{\substack{\sigma \text{ probability }\\\text{ measure on $\R^d$}}}\left\{\frac12\int\int V(|x-y|)\,d\sigma(x)\;d\sigma(y)+\int V_{\rm ext}(x)\,d\sigma(x)\right\}:=a.
\label{eq:1er_ordre}
\end{equation}
The measure $\sigma$, a solution to this variational problem, is in general absolutely continuous with respect to the
Lebesgue measure, and gives the average repartition of the points in space. More specifically,
\begin{equation}
\frac1N\sum_{i=1}^N\delta_{x_i}\wto \sigma
\label{eq:CV_mesure_1er_ordre}
\end{equation}
weakly-$\ast$ in the sense of measures. Since in $\R^d$ the points $x_i$ will have a typical distance of order $N^{-1/d}$, the measure $\sigma$ plays the same role as the macroscopic measure $\rm M$ discussed in~\eqref{eq:def_mu_dilatee}--\eqref{eq:cv_mu}.
For instance, in the case of the Thomson problem for which $V_{\rm ext}$ confines the
particles to the unit sphere $S^2$ and $V(|x|)=|x|^{-1}$, the solution is unique and equal to the uniform measure
$\sigma=(4\pi)^{-1}$ on $S^2$. This means that the particles tend to be uniformly distributed on the sphere. For a
general set $\cM$ or general confining potential $V_{\rm ext}$, proofs of~\eqref{eq:1er_ordre}
and~\eqref{eq:CV_mesure_1er_ordre} are given in~\cite{Choquet-58,Landkof-72,MesSpo-82,CagLioMarPul-92,CagLioMarPul-95,Kiessling-93,Kiessling-09c,Rougerie-15}.

A change of scale is needed to be able to study accurately how the particles are organized at the microscopic scale. In order to do so, it is
better if the potential $V$ behaves appropriately under dilations. In general, one assumes that $V(|x|)=|x|^{-s}$ (or
$V(|x|)=-\log|x|$, which formally corresponds to the case $s=0$). When $V_{\rm ext}$ is smooth, after a dilation of
$N^{-1/d}$ around a given point $\bar x\in\R^d$, the problem happens to coincide with Wigner's model in dimension
$d$, with the local density $\rho=\sigma(\bar x)$. The total limit energy is the superposition of these local
problems, and one finds
\begin{equation}
E_{V_{\rm ext}}(N)=aN+N^{\frac{s}{d}}\,e_{\rm Jell}(1)\int_{\R^d}\sigma(x)^{1+\frac{s}{d}}\,dx+o(N^{\frac{s}{d}}),
\label{eq:expansion_trap_Coulomb}
\end{equation}
where $a$ is the constant given by~\eqref{eq:1er_ordre}, and where $e_{\rm Jell}(1)$ is the Jellium
energy~\eqref{eq:limit_Coulomb} for $\rho=1$ with interaction $V(x)=|x|^{-s}$ (this expansion is modified in the
case $V(x)=-\log|x|$). This result has been recently proved by~\cite{SanSer-14,SanSer-14a,RotSer-14} in the case
$V(x)=-\log|x|$ in dimensions $d=1,2$, in~\cite{RouSer-14} for the Coulomb potential $s=d-2$, and in~\cite{PetSer-14} for $d-2<s<d$.

In the case of a sub-manifold $\cM\subset \R^d$, the scaling is modified, and instead one applies a dilation of $N^{-1/d'}$
where $d'$ is the dimension of $\cM$. It is expected that the same kind of results
hold~\cite{Wagner-90,Wagner-92,GlaEve-92,RakSafZho-94,Brauchart-06,BorHarSaf-07,BraHarSaf-12}, although it has not been proved
yet, except in the case of the sphere in dimension $d=2$ with $V(x)=-\log|x|$~\cite{Betermin-14}.

The asymptotics $O(N)$ in~\eqref{eq:1er_ordre} is only valid if $V$ is locally integrable, so that the right
side is finite. Several authors have studied the case of a potential which is not locally integrable, typically
$V(x)=|x|^{-s}$ for $s\geq d$. In the case of a submanifold $\cM$ of dimension $d'$, the corresponding energy
behaves like $N^{s/d'}$ (or $N\log N$ for $s=d'$). If $s>d'$, it was proved
in~\cite{KuiSaf-98,GotSaf-01,HarSaf-04,HarSaf-05,MarMayRakSaf-04,Borodachov-12} that the corresponding term reads
\begin{equation}
\lim_{N\to\ii}\frac{E_{V_{\rm ext}}(N)}{N^{s/d'}}=|\cM|^{-s/d'} e(1),
\label{eq:limite_non_loc_integrable}
\end{equation}
where $e(1)$ is now the minimal energy~\eqref{eq:limite_rho_fixee} for the problem on the
whole space with $V(x)=|x|^{-s}$:
$$e(1)=\lim_{\substack{N\to\ii\\ |\Omega_N|\to\ii\\ N/|\Omega_N|\to1}}\left(\frac{1}{N}\inf\left\{\sum_{1\leq
      i<j\leq N}|x_i-x_j|^{-s},\ x_i\in \Omega_N\right\}\right).$$
As we pointed out in Section~\ref{sec:fonctions_speciales}, the conjecture is that the particles are located on a
hexagonal lattice in dimension $d'=2$ and FCC when $d'=3$. In such a case, the right side is equal to
$\zeta_{d'}(S,s)$ with $S$ corresponding to the optimal lattice~\cite{Brauchart-06}.

Except in dimension 1~\cite{SanSer-14a,Leble-14} for which the problem is better
understood, it seems that none of these works provide any new information on the crystallization conjecture itself. Nevertheless, they give
an important insight on how it naturally arises in many different situations.

\subsubsection{Vortices and crystallization in dimension 2}

In dimension $d=2$, the crystallization problem appears when
studying fast rotating Bose-Einstein condensates or superconductors in large magnetic field. Vortices are
created, and their number grows with the rotation speed (or the magnetic field intensity). When this number becomes
large, they seem to form a hexagonal lattice, called \emph{Abrikosov lattice} in this context~\cite{Abrikosov-57}.

\begin{figure}[t]
\centering
\includegraphics[height=6cm]{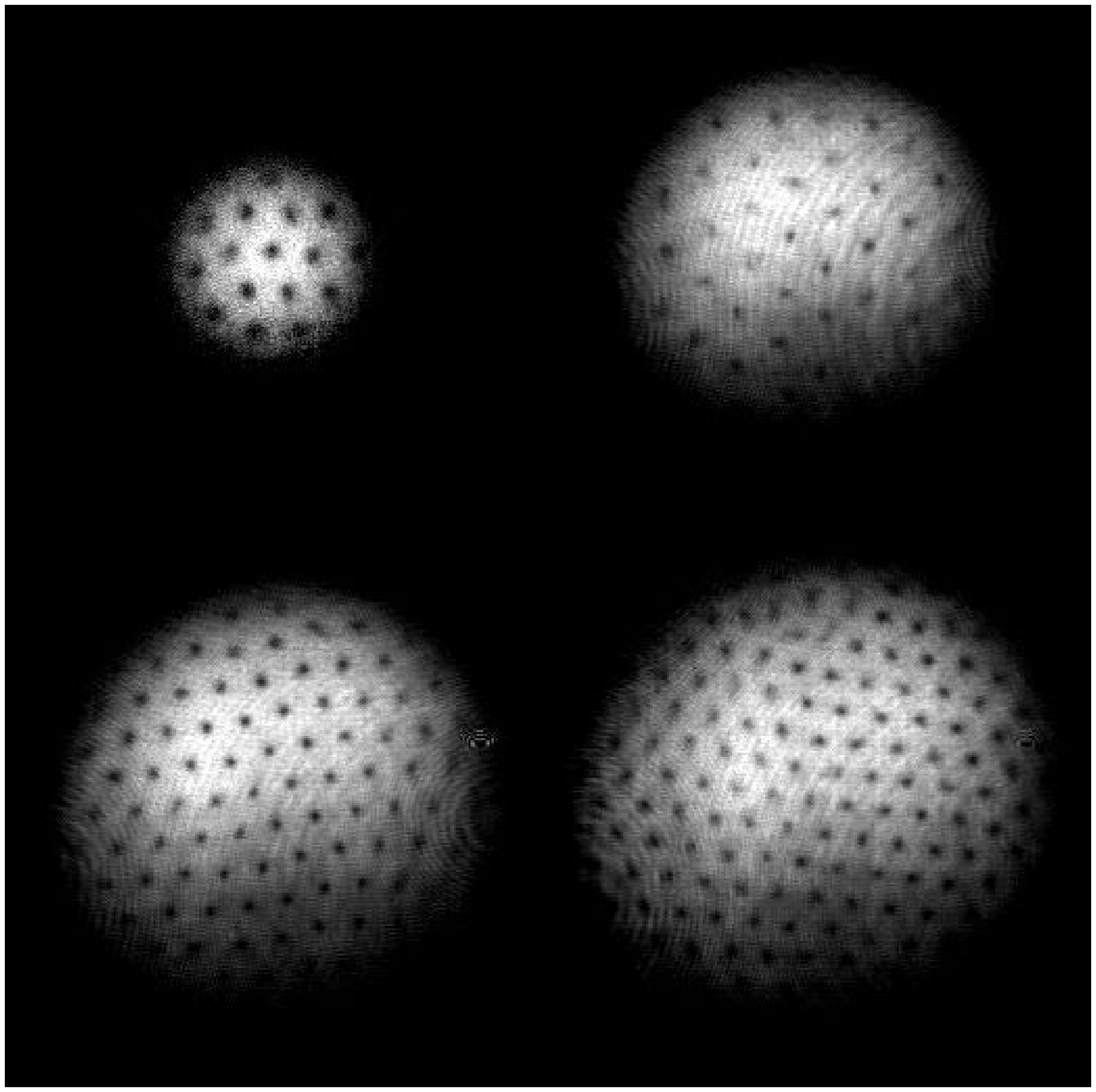}
\includegraphics[height=6cm]{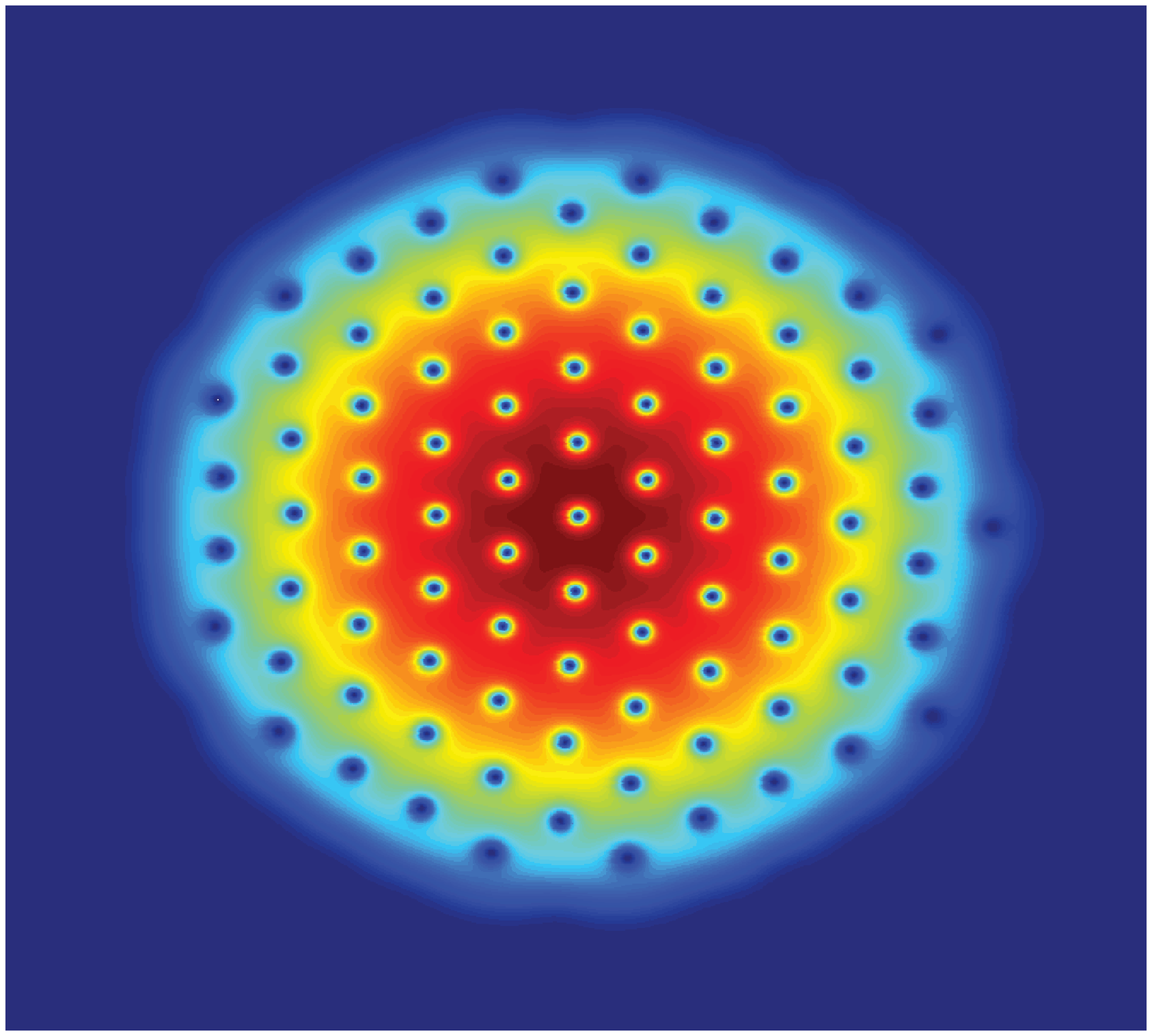}
\caption{\small \emph{Left:} Experimental pictures of fast rotating Bose-Einstein condensates: the number of vortices increases
  with the rotation velocity. The experiments have been conducted by Ketterle's team~\cite{Ketterle-01} at MIT in
  2001. {\scriptsize\copyright\ AAAS.}
  \emph{Right:} Numerical simulation of the Gross-Pitaevskii equation with the software GPELab~\cite{AntDub-14a,AntDub-14b}, reproducing the
  vortex lattice in the corresponding regime.
\label{fig:vortices}}
\end{figure}

In fast rotating Bose-Einstein condensates, vortices may be modeled as classical particles interacting via a potential. The
corresponding energy may be computed using the Jacobi theta
function~\eqref{eq:theta_Jacobi}~\cite{AftBlaNie-06}. In this context, Montgomery's result explains why the
vortices should form a hexagonal lattice (see Figure~\ref{fig:vortices}).

Vortex patterns for the Ginzburg-Landau equation of superconductivity have been widely studied in the mathematics literature
(see~\cite{BetBreHel-94,BetRiv-99}, the first articles on the subject, using simplified models). Under some
constraints on the magnetic field, there is a finite number of vortices which behave like classical particles
interacting via the (two-dimensional) Coulomb potential and submitted to a harmonic confining
potential~\cite{Sandier-98,Jerrard-99,Serfaty-99,Serfaty-99b,Sandier-Serfaty}. For extremely intense magnetic fields, the
number of vortices tends to infinity, and the limit problem becomes that of Wigner's crystallization (see
Section~\ref{sec:Wigner}), as shown in~\cite{SanSer-12}. This explains, although it has not been proved
rigorously yet, why the hexagonal lattice appears in superconductors. We refer to~\cite{Serfaty-14} for a more
detailed presentation of this problem.

\subsubsection{Ohta-Kawasaki model}
\label{sec:otah}

The Ohta-Kawasaki model describes phase separation in copolymer systems~\cite{OhtKaw-86}. In its simplest version, it
consists in minimizing the energy functional
$$\epsilon\, |\partial \{u= 1\}|+\frac12\int_{\Omega}\int_{\Omega}(u(x)-\bar u)V(|x-y|)(u(y)-\bar u)\,dx\,dy$$
where $\Omega$ is a bounded domain of $\R^d$ and where $u$ is allowed to take only the two values $\pm1$, each one
corresponding to a phase of the system. The first term is the perimeter of the set where $u=1$ whereas the second term describes the interaction between the two phases.
The potential $V$ is often assumed to be the Coulomb or Yukawa
interaction, with periodic boundary conditions on $\Omega$ if it happens to be a cube. The question is to determine the
optimal configurations, as the parameters $\varepsilon$ and $\bar u$ vary. If $\bar u=-1$, then the energy
simplifies into
\begin{equation}
  \label{eq:otah-kawasaki}
  \epsilon\, |\partial E|+2\int_{E}\int_{E}V(|x-y|)\,dx\,dy
\end{equation}
with $E=\{u=1\}$.
A regularized version of the model consists in minimizing the energy functional
\begin{equation}
  \int_\Omega \left(\epsilon^2\left|\nabla u(x)\right|^2 + F(u(x))\right)dx  + \int_\Omega \int_{\Omega}(u(x)-\bar u)V(|x-y|)(u(y)-\bar u)\,dx\,dy,
  \label{eq:regularized_Ohta-Kawasaki}
\end{equation}
where $F$ is a non-negative function having as unique minimum points $u=\pm 1$. In the limit $\epsilon\to 0$, this
problem becomes equivalent to \eqref{eq:otah-kawasaki}.

In dimension one, it has been proved that the minimizer is
periodic~\cite{Muller-93,AlbMul-01,RenWei-03,CheOsh-05,Yip-06,GiuLebLie-09} if $\bar u=0$, for all $\epsilon>0$.
Very few results exist in higher dimension~\cite{AlbChoOtt-09,ChoPel-10}.
In the limit where one phase is strongly favored ($\bar u\sim -1$) and $\epsilon\to0$, it has been proved that,
here again, the opposite phase $u=1$ is a solution to Wigner crystallization problem~\cite{GolMurSer-13,GolMurSer-14}.

A proof of crystallization on the hexagonal lattice (in 2D) has been recently given in ~\cite{BouPelThe-14} for a
different copolymer model. In this theory, the second term in  \eqref{eq:otah-kawasaki} is replaced by the
Wasserstein distance $\mathcal{W}$ to the Lebesgue measure. Hence, the energy is defined for point
measures $\mu$ having their support in $\Omega\subset\R^2$. It reads
$$\cE(\mu)=\epsilon \sum_{z\in{\rm supp}(\mu)}\sqrt{\mu(\{z\})}+\mathcal{W}(\1_\Omega,\mu),$$
and crystallization is proved for any $\varepsilon>0$ if $\Omega$ has appropriate symmetries, and if $\varepsilon$
is sufficiently small (or equivalently if $\varepsilon>0$ is fixed and $|\Omega|\to\ii$). In the limit
$\varepsilon\to0$ it had been proved previously that the hexagonal lattice minimizes the Wasserstein distance to
the Lebesgue measure~\cite{Newman-82}.

 \subsection{The macroscopic object and its microscopic structure}\label{sec:mu}

We mentioned above the question of proving the existence of a macroscopic measure ${\rm M}$, obtained as the weak
limit~\eqref{eq:cv_mu}. This is related to the formation of a macroscopic object. Another question is to know what
kind of object is formed, that is, to compute the measure ${\rm M}$.

This problem seems different from the local behavior of the particles. However, the hexagon in Figure~\ref{fig:lj}
indicates that a link exists with the microscopic scale. Indeed, if exact crystallization is assumed, that is, if the
particles are restricted to be on the vertices of a periodic lattice for all $N$, then it is possible to write a
limit minimization problem for the surface energy, which coincides with the second-order term in the development of
$E(N)$. This term is of order $N^{(d-1)/d}$. It has been proposed by Wulff~\cite{Wulff-01}, and proved rigorously
for a hard sphere model\footnote{That is,
  $V\equiv+\infty$ sur $[0,1-\epsilon)$, $V\equiv0$ on
  $[1+\epsilon,\infty)$ and $\min(V)=V(1)<0$, with small $\epsilon$.} in dimension two by Au Yeung, Friesecke and Schmidt
in~\cite{AuFriSch-12,Schmidt-13}. This work is based on results by
Radin et al~\cite{Radin-81,HeiRad-80}. We refer for instance to
\cite{BodIofVel-00,Bodineau-99,CerPis-00} for similar results on the Ising model.

These works \emph{assume} that the particles form a subset of a given periodic lattice for all $N$, which is true
only for very specific interaction potentials $V$. It would be interesting to generalize these results to more
general cases. However, this problem is \emph{a priori} a very difficult one, since a good knowledge of the leading
order term of $E(N)$ is needed to understand the next one. And this is exactly the crystallization conjecture.

\section{Extensions}
\label{sec:extensions}

\subsection{Positive temperature}
\label{sec:temperature}

Until now we have only considered the problem of minimizing the energy, that is, we have assumed the temperature to
be $0$. As a matter of fact, it seems intuitive that crystallization only occurs for small
temperature~\cite{Radin-87}. At positive temperature $T>0$, the problem is more complicated to state. We discuss here a possible formulation and mention some important references.

The point particles should now be replaced by a probability density on $\R^{dN}$ that only describes random positions of the $N$ particles. This probability is found by minimizing the free energy, which is the sum of the energy $\cE_N$ and of an entropy term multiplied by the temperature $T$. The entropy favors extended systems and a high value of $T$ will spread out the probability density. In the whole space the entropy would make all the particles fly apart, and one therefore needs to confine them. This may be done as in Section~\ref{sec:densite_fixe} by imposing that the system is in a bounded domain $\Omega$ and that the volume $|\Omega|$
is proportional to the (average) number of particles $N$. Another possibility would be to work in the whole space $\R^d$, and add a confining
external potential as was done in~\eqref{eq:riesz}, but we will (almost) not discuss this here.

\subsubsection*{Formulation in the canonical ensemble}

The probability that we have to consider is the Gibbs probability measure\footnote{It is also possible to consider the
Hamiltonian~\eqref{eq:Hamiltonien} instead of $\cE_N$, but the Gibbs measure $e^{-\cH_N/T}$ can then be factorized and the variables
$p_i$ do not play any role. The situation is different in the quantum case (Section~\ref{sec:quantique}).}
\begin{equation}
P_{\Omega,N,T}(x_1,...,x_N)=\frac{\displaystyle e^{-\frac{\cE_N(x_1,...,x_N)}{T}}}{\displaystyle\int_{\Omega^N}e^{-\frac{\cE_N(x_1,...,x_N)}{T}}\,dx_1\cdots dx_N}.
 \label{eq:Gibbs}
\end{equation}
This distribution concentrates on the minima of $\cE_N$ when $T\to 0$. This probability measure is obtained by minimizing the
\emph{free energy}
\begin{multline}
F_{\Omega}(N,T):=\min_{\substack{P' \text{ symmetric measure}\\ \text{on $\Omega^N$, with}\\ \text{$\frac{1}{N!}\int_{\Omega^N} P'=1$}}}\bigg\{\frac{1}{N!}\int_{\Omega^N} \cE_N(x_1,...,x_N)\, P'(x_1,...,x_N)\,dx_1\cdots dx_N\\+\frac{T}{N!}\int_{\Omega^N} P'(x_1,...,x_N)\log P'(x_1,...,x_N) \,dx_1\cdots dx_N\bigg\},
\label{eq:energie_libre}
\end{multline}
in which the first term is the energy of the system, and the second one is the opposite of the entropy. The
symmetry of $P'$ accounts for the fact that the particles are identical and indistinguishable. The Boltzmann coefficient $1/N!$ appearing in front of all the integrals is here to count each possible configuration only once. In~\eqref{eq:energie_libre} we can replace $P'$ by the probability measure $P=P'/(N!)$ at the expense of an additive term $T\log(N!)$. At $T=0$ an optimal $P$ is a delta measure at one minimum of $\cE_N$. But when $T>0$, the entropy term imposes that $P$ be absolutely continuous with respect to the Lebesgue measure.

The solution of the minimization problem~\eqref{eq:energie_libre} is unique, given by $P'=(N!)P_{\Omega,N,T}$ in~\eqref{eq:Gibbs}, and satisfies
$$F_{\Omega}(N,T)=-T\log Z_\Omega(N,T),$$
where
$$Z_\Omega(N,T)=\frac{1}{N!}\int_{\Omega^N}e^{-\cE_N(x_1,...,x_N)/T}\,dx_1\cdots dx_N$$
is called the \emph{(canonical) partition function}. As before we look at the coupled limit $N\to\ii$ with $|\Omega_N|\to\ii$ and $N/|\Omega_N|\to\rho>0$, a fixed parameter.  Using the stability of $V$, $\cE_N\geq N\,e_\ii$, we find
$$F_{\Omega_N}(N,T)\geq -T\log\left(\frac{|\Omega_N|^N}{N!}e^{-e_\ii N/T}\right)\sim N\big(e_\ii+T(\log\rho-1)\big)$$
by Stirling's formula. Similarly, by Jensen's inequality,
$$F_{\Omega_N}(N,T)\leq -T\log\left(\frac{|\Omega_N|^N}{N!}e^{-|\Omega_N|^{-N}\int_{\Omega_N^N}\cE_N/T}\right)\sim NT\left(\log\rho-1\right)+\frac{N\rho}{2}\int_{\R^d}V(|x|)\,dx.$$
A similar estimate exists when $V$ is not integrable at $0$. We deduce that $F_{\Omega_N}(N,T)$ behaves linearly in $N$. After a little more work~\cite{Ruelle}, one can actually prove that the thermodynamic limit exists
$$\boxed{f(\rho,T)=\lim_{\substack{N\to\ii\\ |\Omega_N|\to\ii\\ N/|\Omega_N|\to\rho}}\frac{F_{\Omega_N}(N,T)}{N},}$$
as in~\eqref{eq:limite_rho_fixee}.

In order to formalize the crystallization problem at positive temperature, it is convenient to consider the \emph{empirical measures} (also called $k$-point correlation functions~\cite{Fisher-65,CarAsh-97}), which are similar to the measure
$\mu_N$ introduced in~\eqref{eq:def_mu_N}. To be more precise, we define the family of measures,
obtained by integrating with respect to all variables except $k$ of them:
\begin{equation}
\mu_{\Omega,N,T}^{(k)}(x_1,...,x_k)=\frac{N!}{(N-k)!}\int_{\Omega^{N-k+1}}P_{\Omega,N,T}(x_1,...,x_k,y_{k+1},...,y_N)\,dy_{k+1}\cdots dy_N.
\end{equation}
A natural definition of crystallization is that the measures $\mu_{\Omega_N,N,T}^{(k)}$ locally converge, after extraction of a subsequence $N_j$ and applying an appropriate translation $\tau_j$ to some locally finite measures in the thermodynamic limit,
$$\mu_{\Omega_{N_j},N_j,T}^{(k)}(\cdot-\tau_j)\wto \mu_{\rho,T}^{(k)}$$
where these measures are invariant under the action of a (maximal) lattice group $G$:
\begin{equation}
\forall g\in G,\qquad \mu_{\rho,T}^{(k)}(x_1+g,...,x_k+g)=\mu_{\rho,T}^{(k)}(x_1,...,x_k).
 \label{eq:def_cristallisation_positive_temp}
\end{equation}
Here again the limit will depend on the subsequence and on the sequence of domains $(\Omega_N)$. Different weak limits might have different invariance groups. The group $G$ in~\eqref{eq:def_cristallisation_positive_temp} is assumed to leave invariant \emph{all} the possible weak limits up to rotations (for chosen $T$ and $\rho$ and all $N_j$ and $\Omega_{N_j}$ for which the limit exists).

In general, when $\mu^{(2)}(x_1,x_2) - \mu^{(1)}(x_1)\mu^{(1)}(x_2)$ does not tend to zero when $|x_1-x_2|\to\ii$, one says that there is \emph{long range order}, meaning that two particles far away are correlated, although the system is not necessarily periodic. Many works have been devoted to the proof that systems exhibit long range order, without reaching exact periodicity.

For physical systems one might expect that there will be a periodic (or at least long-range) order for all $T\leq T_c(\rho)$ for a small critical temperature $T_c(\rho)$, and that the system will be translation-invariant, with unique equilibrium states, for $T\geq T_c'(\rho)\geq T_c(\rho)$. It is usually believed that the function $f(\rho,T)$ is piecewise real-analytic in $(T,\rho)$ and that the curves of non-analyticity correspond to phase transitions~\cite{Ruelle}.

In the zero-temperature case, we only considered the measure $\mu^{(1)}$, which appears
in~\eqref{eq:cv_mu_cristal}. The reason is that, in such a case, a solution to~\eqref{eq:energie_libre} is
$$P_{\Omega,N}(y_1,...,y_N)=\frac1{N!}\sum_{\sigma\in\mathfrak{S}^N}\delta_{x_{\sigma(1)}}(y_1)\cdots \delta_{x_{\sigma(N)}}(y_N)$$
where $x_1, \dots ,x_N$ is a solution to problem~\eqref{eq:minimisation}. Because of this specific form, local
convergence for $k=1$ to $\mu$ implies that all the other empirical measures $\mu^{(k)}(x_1,...,x_k)$ automatically converge
to $\mu(x_1)\cdots \mu(x_k)$. Such a property cannot hold in general for $T>0$. In case of crystallization at $T=0$, it is usually expected that the $k$-particle densities $\mu_{\rho,T}^{(k)}$ should converge in the limit $T\to0$ towards the \emph{uniform average} of translations and rotations of the crystal:
$$\lim_{T\to0}\mu^{(k)}_{\rho,T}(x_1,...,x_k) = \frac{1}{|Q|}\int_Q\int_{SO(d)}\prod_{j=1}^k \mu\big(R(x_j -\tau)\big)\, {\rm d}R\, {\rm d}\tau:=\mu_{\rho,0}^{(k)}(x_1,...,x_k).$$
Then $\mu_{\rho,0}^{(k)}$ is translation-invariant (in particular, $\mu^{(1)}_{\rho,0}$ is constant), but $\mu_{\rho,0}^{(2)}$ is periodic in $x_1-x_2$.

\subsubsection*{Grand canonical ensemble}
The problem is often stated in the grand-canonical ensemble, for which the algebra is simpler. This corresponds to assuming that the number of particles $N$ in $\Omega$ is also a random variable. It is then customary to consider the Laplace transform of the measures $N\mapsto Z_\Omega(N,T)$, introducing a variable $\mu=(T/N)\log z$ dual to $N$, called the \emph{chemical potential}:
$$\tilde Z_\Omega(z,T):=\sum_{N\geq0}e^{\mu N/T}Z_\Omega(N,T)=1+\sum_{N\geq1}\frac{z^N}{N!}\int_{\Omega^N}e^{-\cE_N/T}.$$
The series on the right is conveniently written in terms of the \emph{fugacity} $z=e^{\mu /T}$. The stability $\cE_N\geq e_\ii N$ makes the sum convergent, with $\tilde Z_\Omega(z,T)\leq \exp(|\Omega|e^{(\mu-e_\ii)/T})$.
By varying $z>0$ (or $\mu$), one can obtain some information on the canonical case. The problem is then the same as before, except that we cannot divide by $N$ which is an unknown variable, so we instead divide by $|\Omega|$ and define
$$ \tilde f(z,T)=\lim_{|\Omega|\to\ii}\frac{-T\log \tilde Z_\Omega(z,T)}{|\Omega|},$$
which now satisfies $-T |\Omega|e^{(\mu-e_\ii)/T}\leq \tilde f(z,T)\leq0$.
In a similar manner, we define the $k$-particle grand-canonical correlation function by
\begin{multline}
\tilde\mu_{\Omega,z,T}^{(k)}(x_1,...,x_k)\\=\tilde Z_\Omega(z,T)^{-1}\left(z^ke^{-\frac{\cE(x_1,...,x_k)}T}+\sum_{n\geq1}\frac{z^{k+n}}{n!}\int_{\Omega^n}e^{-\frac{\cE(x_1,...,x_{k+n})}T}dx_{k+1}\cdots dx_{k+n}\right)
\label{eq:def_correlation_fn_grand_canonique}
\end{multline}
and ask the same questions as before about its possible local weak limits.

\subsubsection*{Known results}

There are many results showing the \emph{absence of crystallization} in several situations, but almost none proving its existence. The first classical theorems are due to Ruelle and Penrose~\cite{Ruelle,Ruelle-63,Penrose-63} who studied the convergence of the series at small fugacity $z$ (hence, high-temperature for fixed $\mu$). They showed that the grand-canonical free energy as well as the correlation functions are all convergent series in the parameter $z=e^{\mu /T}$, with a radius of convergence at least equal to
\begin{equation}
R_{\rm min}(T)=e^{2e_\ii/T-1}\left(\int_{\R^d}\left|e^{-V(x)/T}-1\right|\,dx\right)^{-1}.
\label{eq:radius_CV}
\end{equation}
Furthermore, the correlation functions must be translation-invariant when $z$ and $T$ are in this range. Since $e_\ii\leq0$ with strict inequality if $\min V<0$, the radius $R_{\rm min}(T)$ shrinks exponentially fast when $T\to0$. The main tool here is a system of equations for the correlation functions called the \emph{Kirkwood-Salsburg equations}, that allows to derive uniform bounds of the form $\mu^{(k)}(x_1,...,x_k)\leq C^k$ at small fugacity. The bound on the radius of convergence was recently improved in~\cite{ProYuh-15}. The series of $\tilde f(z,T)$ in terms of $z$ is called the \emph{Mayer expansion}. It is possible to invert the value of the constant density $\rho(z,T)=\tilde \mu^{(1)}$ in terms of $z$ and to express $\tilde f$ as a function of $\rho$ and $T$. Since $\rho(z,T)=z+O(z^2)$, this is a low-density regime. The corresponding series (called the \emph{Virial expansion}) was studied in~\cite{LebPen-64}.

The properties of the system in the thermodynamic limit depend in a crucial way on the space dimension $d$, a
phenomenon that is not present at $T=0$. For one-dimensional systems with a fastly-decaying potential, a result due
to van Hove in~\cite{vanHove-50} (see also~\cite[Thm~5.6.7]{Ruelle} and~\cite{Ruelle-68,Dobrushin-68}) states that
$\tilde f(z,T)$ is actually real-analytic for all $z>0$ with unique correlation functions, hence there is no phase
transition. Given that crystallization has been proved for a large variety of models in dimension one at zero
temperature, this shows that the phase transition occurs exactly at $T_c(\rho)=0$. We notice that phase transitions
can occur for an integrable potential decaying as slowly as $1/|x|^2$~\cite{Dyson-69,FroSpe-82}. A similar result
called the Mermin-Wagner theorem~\cite{MerWag-66,Mermin-67,Mermin-68,FroPfi-81,FroPfi-86} exists in two dimensions
for very short-range potentials, except that only the translation-invariance is known, and there does not seem to be any information on the range of analyticity.

Given that there is no crystalline order in one and two-dimensional systems at positive temperature, but that crystallization is expected at $T=0$, a natural question is to ask what is happening when $T\to0$. Recent works~\cite{ColKonMorSid-10,Jansen-12,JanKon-12} investigate the exponentially small region of convergence of the Mayer series close to $T=\rho=0$, where things can be proved in any dimension. Namely, they considered the limit $T\to 0$ and $\rho\to0$ with the constraint that $T\log\rho\to \nu$ (recall that $\rho\sim z=e^{\mu/T}$ at small activity, so this is similar to fixing $\mu$). They showed that, in any dimension, the system behaves as a gaz of finite clusters of particles which essentially do not interact. The number of particles $K$ in the clusters is determined by the parameter $\nu$ and it can be as big as we want if all the $E(K)$ have minimizers. Each cluster of $K$ particles solves the problem $E(K)$. For $K\to\ii$ the cluster converges to the zero-temperature crystal configuration.

Even without more information about the limiting states, it is possible to show in most situations that the correlation functions satisfy a pointwise bound of the form
\begin{equation}
\tilde \mu_{\Omega,z,T}^{(k)}(x_1,...,x_k)\leq (C_{z,T})^k
\label{eq:uniform_bd_correlation}
\end{equation}
and therefore have local weak-$\ast$ limits, after extraction of subsequences. The positive temperature makes the weak convergence of the correlation functions much easier than at $T=0$ (see Section~\ref{sec:CV_mu_N}), but the constant $C_{z,T}$ always blows up when $T\to0$.
The first simple case is that of a non-negative potential $V$: using in~\eqref{eq:def_correlation_fn_grand_canonique} that $\cE_N(x_1,...,x_{k+n})\geq \cE_N(x_{k+1},...,x_{k+n})$ we immediately obtain the pointwise bound~\eqref{eq:uniform_bd_correlation} with $C_{z,T}=z$.
The most general case was covered by Ruelle in~\cite{Ruelle-70} who proved~\eqref{eq:uniform_bd_correlation} for \emph{super-stable interactions $V$}. Super-stability means that there exists constants $A,B,r>0$ such that
\begin{equation}
 \sum_{1\leq i<j\leq N}V(|x_i-x_j|)\geq \sum_{k\in \Z^d}\Big(A\,n(k)^2-B\,n(k)\Big)
 \label{eq:superstable}
\end{equation}
where $n(k)$ is the number of particles in the cube $kr+[-r/2,r/2)^d$. Any continuous non-negative interaction with $V(0)>0$ satisfies~\eqref{eq:superstable} for $r$ small enough. Hence any stable interaction can be made super-stable by slightly increasing its values in a neighborhood of $0$. Ruelle's constant $C_{z,T}$ is, however, a complicated function of the parameters. It was shown in~\cite{FroPar-78} that one can take $C_{z,T}=ze^{w(0)/T}$ when $\widehat{w}\geq0$.

The Jellium problem described in Section~\ref{sec:Wigner} can be defined at positive temperature and it is an exception for which crystallization
occurs for any $T$ and any density $\rho$ in dimension $d=1$~\cite{Kunz-74,BraLie-75,AizMar-80}. This is not surprising, since screening effects should make the effective potential be integrable at infinity, but probably not decay faster than $1/r^2$.
Numerical simulations indicate that, in dimension~$d\geq2$, there exists a
critical temperature $T_c>0$ such that, if $T>T_c$, then Jellium is not crystallized~\cite{BruSahTel-66,Hansen-73,PolHan-73,GanChaChe-79,AlaJan-81,LeePerrSmi-82,Alastuey-86,Stishov-98,DubNei-99,BryMar-99,Bonetal-08}. The link between a trapped Coulomb gas in the mean-field limit and the positive-temperature Jellium problem was recently studied in~\cite{LebSer-15,Leble-15}, similarly to what we discussed in~\eqref{eq:expansion_trap_Coulomb}.

\subsubsection*{Link with random matrices}

The $N$ eigenvalues of an $N\times N$ matrix with random coefficients are, in some situation, distributed
according to the Gibbs measure of a gas of particles in an external trapping potential. The effective interaction will usually be $V(x)=-\log|x|$, and the dimension
$d=1$ (if the eigenvalues are real) or $d=2$ (if they are complex). If the entries of
the matrix are independent Gaussian variables, the statistical distribution of the eigenvalues
$\lambda_1,...,\lambda_N$ is given by the Gibbs measure~\eqref{eq:Gibbs} with the mean-field energy
$$\cE_N(\lambda_1,...,\lambda_N)=-\frac 1 N\sum_{1\leq k<\ell\leq N}\log|\lambda_k-\lambda_\ell|+\sum_{j=1}^N|\lambda_j|^2.$$
For hermitian matrices (\emph{GUE}, that is, \emph{Gaussian Unitary Ensemble}), the problem is set in
$\Omega = \R$, since the eigenvalues have no imaginary part. In such a case, the temperature is equal to
$T=1/(2N)$. If one imposes that the matrices have real coefficients (\emph{GOE}, that is,
\emph{Gaussian Orthogonal Ensemble}), the temperature is $T=1/N$. When considering complex matrices without any
symmetry assumption (\emph{Ginibre ensemble}), we have the same formula, but the $\lambda_i$ are now in
$\Omega=\C=\R^2$ and the temperature is $T=1/(2N)$.

It is also possible to consider unitary or orthogonal matrices (\emph{CUE} for \emph{Circular Unitary Ensemble}, and
\emph{COE} for \emph{Circular Orthogonal Ensemble}, respectively),
using the uniform law on this compact subset of matrices. Then, the eigenvalues are distributed according to the
Gibbs measure~\eqref{eq:Gibbs} with
$$\cE_N(\lambda_1,...,\lambda_N)=-\frac 1 N\sum_{1\leq k<\ell\leq N}\log|\lambda_k-\lambda_\ell|,$$
this time restricted to the unit circle $\Omega=S^1$.

Studying the eigenvalues of random matrices and the link with Coulomb gas is a very active subject, which started with
the seminal works of Wigner~\cite{Wigner-55,Wigner-67} and
Dyson~\cite{Dyson-62a,Dyson-62b,Dyson-62c,DysMeh-63a,DysMeh-63b}. The interest in the set of matrices we just
mentioned is that they allow for explicit computation of empirical measures, hence a good knowledge of the
statistics of theses eigenvalues. Since $T$ behaves like $1/N$, the first order corresponds to the zero-temperature
setting. The average distribution of the eigenvalues is given by the measure $\sigma$ solution to the
minimization problem in~\eqref{eq:1er_ordre}. The next order is more complex and its link
with the crystal problem is less clear~\cite{Dyson-62c}. We refer for instance
to~\cite{Mehta-10,Forrester-10,AndGuiZei-10} for a detailed study of the subject.

\subsection{Several types of particles}\label{sec:several_types}

In order to deal with long-range interactions (for instance Coulomb potential), as in Wigner problem presented in
Section~\ref{sec:Wigner}, it is possible to add a background homogeneous density making the system globally
neutral. Another model, more important from a practical viewpoint, is the case of two (or more) different types of
atoms or ions, with different charges. One can think for instance of sodium chloride crystal, which is made of two
face centered cubic lattices, one (Na$^+$ ions) shifted with respect to the other (Cl$^-$ ions).

For the sake of simplicity, let us consider only two types of particles. The interaction between two identical
particles is different from the interaction between two different ones. We are thus led to the energy
\begin{multline*}
\cE_{N_1,N_2} \left(x_1,\dots,x_{N_1},y_1,...,y_{N_2}\right) \\= \sum_{1\leq i < j\leq N_1}
V_{11}\left(\left|x_i- x_j\right|\right)+\sum_{1\leq i < j\leq N_2}
V_{22}\left(\left|y_i- y_j\right|\right)+\sum_{i=1}^{N_1}\sum_{j=1}^{N_2} V_{12}\left(\left|x_i- y_j\right|\right)
\end{multline*}
where $x_i$ and $y_i$ are the positions of the particles of each type. We study the limit $N_1,N_2\to\ii$,
possibly imposing a link between $N_1$ and $N_2$, accounting for a charge difference between the two types of
particles. Thinking of a 3D crystal composed of charges of opposite sign $q_1$ et $-q_2$, we assume that
$$V_{11}(|x|)\mathop{\sim}_{|x|\to\ii}\frac{q_1^2}{|x|},\qquad
V_{22}(|x|)\mathop{\sim}_{|x|\to\ii}\frac{q_2^2}{|x|},\qquad
V_{12}(|x|)\mathop{\sim}_{|x|\to\ii}-\frac{q_1q_2}{|x|},$$
and we impose the neutrality condition $q_1N_1-q_2N_2\to0$ in the limit. For such a classical model, the Coulomb
interaction is not adapted, since the energy tends to $-\ii$ as two particles of opposite charge get closer to
each other, and the model is unstable (note, however, that it is stable in the quantum case, as it was proved by
Dyson-Lenard~\cite{DysLen-67,DysLen-68} and Lieb-Thirring~\cite{LieThi-75}). Hence, one needs to assume that the
potentials $V_{11}$ $V_{22}$ and $V_{12}$ are repulsive at short distance $|x|\to0$. The simplest choice is to take
$$V_{ij}(|x|)=2(\delta_{ij}-1/2)q_iq_j W(x),\quad \text{with}\quad W(x)=\frac{1-e^{-\mu|x|}}{|x|}.$$
Since $W$ has a positive Fourier transform, the same calculation as in~\eqref{eq:argument_positive_Fourier} gives that the total interaction is bounded from below by
\begin{multline*}
\cE_{N_1,N_2} \left(x_1,\dots,x_{N_1},y_1,...,y_{N_2}\right)\\
\geq \frac1{2(2\pi)^{d/2}}\int_{\R^d}\widehat{W}(k)\left|q_1\sum_{j=1}^{N_1}e^{ik\cdot x_j}-q_2\sum_{k=1}^{N_2}e^{ik\cdot y_k}\right|^2\,dk-\frac{N_1q_1^2+N_2q_2^2}{2}\mu\geq -\frac{N_1q_1^2+N_2q_2^2}{2}\mu
\end{multline*}
and the model is stable.

Several conjectures have been made concerning the optimal lattices~\cite{BorGlaMPh-13}, but we do not know any
result on the crystal problem with several types of particles. Thinking of crystalline structures currently observed
in nature, it is a highly important question from a physical viewpoint. The existence of the thermodynamic limit and estimates on the correlation functions were proved in~\cite{FroPar-78}. A review of known results in 3D for high
temperature (hence without crystallization) is given in  \cite{BryMar-99}. In~\cite{Radin-86} Radin considers special
short-range potentials for two types of particles, and proves that crystallization fails, but the minimizers are
quasi-periodic.

\subsection{Quantum models}\label{sec:quantique}

In the classical models studied so far, the kinetic energy of the particles does not play any role, since we
deal with minimizers or Gibbs states. The term
$\sum_{j=1}^N|p_j|^2/(2m)$ in~\eqref{eq:Hamiltonien} disappears in the minimization problem, and factors out and
gives a Gaussian at positive temperature in~\eqref{eq:Gibbs}. The situation is different in quantum mechanics, in
which there is a link between velocity and position, in order to respect Heisenberg's uncertainty principle. This
makes the kinetic energy dependent on the positions of the particles. More precisely, quantum mechanics principles
imply that $p_j$ should be replaced by the differential operator $-i\hbar\nabla_{x_j}$ and that the Hamiltonian
$\cH_N(p_1,...,p_N,x_1,...,x_N)$ in~\eqref{eq:Hamiltonien} should be replaced by the differential operator
\begin{equation}
H_N=-\sum_{j=1}^N\frac{\hbar^2}{2m}\Delta_{x_j}+\sum_{1\leq k<\ell\leq N}V(x_k-x_\ell).
\end{equation}
This operator acts on $L^2(\Omega^N)$, where $\Omega=\R^d$ for an unconfined system, and where $\Omega$ is a
bounded domain if the system is confined (with suitable boundary conditions). Since the particles are
indistinguishable, we work with a subspace of $L^2(\Omega^N)$ consisting of functions having a prescribed symmetry
property. In nature one can find two types of particles: \emph{bosons} and \emph{fermions}. For bosons, we use the subspace
$L^2_s(\Omega^N)$ of functions which are symmetric with respect to variable permutations. For fermions, we use the
subspace $L^2_a(\Omega^N)$ of functions which are antisymmetric. Properties of the system in the limit $N\to\ii$
depend on the chosen symmetry class. For the sake of simplicity, we asume that the particles have no spin.

The classical problems studied so far read, in the quantum case,
\begin{multline}
E_{a/s,\Omega}(N)\\=\inf_{\substack{\Psi\in L^2_{a/s}(\Omega^N)\\ \int|\Psi|^2=1}}\int_{\Omega^N}\left(\frac{\hbar^2}{2m}|\nabla\Psi(x_1,...,x_N)|^2+\sum_{1\leq k<\ell\leq N}V(x_k-x_\ell)|\Psi(x_1,...,x_N)|^2\right)dx_1\cdots dx_N,
\label{eq:min_quantique_Omega}
\end{multline}
for the energy at $T=0$,
$$F_{a/s,\Omega,T}(N)=-T\log\left({\rm tr}_{L^2_{a/s}(\Omega^N)}e^{-H_N/T}\right)$$
for the canonical free energy, and
$$\tilde F_{a/s,\Omega,z,T}(N)=-T\log\left(\sum_{N\geq0}z^N\,{\rm tr}_{L^2_{a/s}(\Omega^N)}e^{-H_N/T}\right)$$
for the grand-canonical free energy. Note the absence of the Boltzmann factor, due to the restriction that we work in the subspaces $L^2_{a/s}(\Omega^N)$. If we instead work in the full space $L^2(\Omega^N)$ (indistinguishable particles sometimes called ``boltzons''), then we need to put back the $1/N!$ coefficient. In the semi-classical limit $\hbar\to0$, these (free) energies converge (up
to a constant which diverges like $\log\hbar$ at positive temperature) to the corresponding classical (free) energies.

Quantum mechanics is by nature a probabilistic theory and the study of crystallization relies on the weak limit of the
empirical measures, as in the case of the positive temperature classical model. For instance, at zero temperature one can study the limit
of the $k$-particle densities
$$\mu^{(k)}_{\Omega,N}(x_1,...,x_k):=\frac{N!}{(N-k)!}\int_{\Omega^{N-k}} |\Psi_{\Omega,N}(x_1,...,x_k,y_{k+1},...,y_N)|^2\,dy_{k+1}\cdots dy_N$$
where $\Psi_{\Omega,N}$ is a minimizer of problem~\eqref{eq:min_quantique_Omega} (this minimizer is always unique,
up to a phase, for bosons, but it is not necessarily unique for fermions). A similar formula is valid at positive temperature but we do not provide the details. The measure $\mu^{(k)}_{\Omega,N}$
corresponds to the one defined in classical mechanics and it does not carry all the information on the quantum system. It is more relevant to study the
limit of  \emph{$k$-particle density operators}, which are defined by their integral kernel
\begin{multline}
\gamma^{(k)}_{\Omega,N}(x_1,...,x_k,x_1',...,x_k')\\:=\frac{N!}{(N-k)!}\int_{\Omega^{N-k}} \Psi_{\Omega,N}(x_1,...,x_k,y_{k+1},...,y_N)\overline{\Psi_{\Omega,N}(x'_1,...,x'_k,y_{k+1},...,y_N)}\,dy_{k+1}\cdots dy_N,
\end{multline}
and where the diagonal part coincides with $\mu^{(k)}_{\Omega,N}$. We say that the system crystallizes if theses operators
locally converge to operators $\gamma^{(k)}$ which commute with translations of a (maximal) lattice $G$, that is,
\begin{displaymath}
  \forall g\in G, \quad
   \gamma^{(k)}_{\Omega,N}(x_1+g,...,x_k+g,x_1'+g,...,x_k'+g) = \gamma^{(k)}_{\Omega,N}(x_1,...,x_k,x_1',...,x_k').
\end{displaymath}

If $V$ is a stable potential, then the quantum mechanical problem is also stable since the Laplacian is a non-negative operator, and the thermodynamic limit can be shown to exist~\cite{Ruelle}. However, potentials $V$ that are unstable classically can become stable in quantum mechanics. This the case for the Coulomb potential in dimension $d=3$ with two kinds of particles of opposite charge which has been shown to be stable for fermions~\cite{LieThi-75,LieThi-76,LieSei-09} but not for bosons~\cite{Dyson-67,ConLieYau-88,LieSol-01,LieSol-04,Solovej-06}. The existence of the thermodynamic limit in the Coulomb case is proved in various situations~\cite{LieLeb-72,FroPar-80,Fefferman-85,HaiLewSol_1-09,HaiLewSol_2-09,BlaLew-12}. Correlation functions were studied in this case in~\cite{FroPar-78,FroPar-80}.

Few results have been proved for the crystal problem in the case of continuum quantum systems. In particular, one
could think that, when a classical system exhibits crystallization, so does the quantum corresponding system if
$\hbar$ is sufficiently small. This has not been studied, to our knowledge, except in the case of Coulomb gas
(quantum Jellium) for which Kunz~\cite{Kunz-74} and Brascamp-Lieb~\cite{BraLie-75} have proved crystallization for
small density $\rho$ in dimension $d=1$. After a change of scale, assuming that $\rho$ is small is equivalent to
assuming $\hbar$ is small, so the situation is indeed a semi-classical limit. Crystallization for 1D
quantum Jellium (at any density $\rho$ and any temperature $T$) has been recently proved by Jansen and Jung~\cite{JanJun-14}.

Many of the results on the non-existence of crystals in the classical case have been extended to quantum systems. Using the Feynmann-Ka\v{c} formula (expressing all the quantum objects in terms of paths in the classical problem), Ginibre proved in~\cite{Ginibre-65a,Ginibre-65b,Ginibre-65c} that the grand-canonical free energy as well as the $k$-particle density matrices are convergent series in terms of the fugacity $z$. Hence there is no phase transition for $z$ small enough and the $k$-particle density matrices are translation-invariant. The uniform bounds~\eqref{eq:uniform_bd_correlation} allowing to define local weak limits have been generalized to quantum systems in~\cite{FroPar-78,FroPar-80,EspNicPul-82,Park-85}.

Some results have been proved for quantum systems described by nonlinear models, such as Thomas-Fermi or
Hartree. In the Coulomb case, assuming that the nuclei are classical particles with positive charge and are distributed on a lattice, it
has been proved for convex models that the electrons are periodically
arranged~\cite{LieSim-77b,CatBriLio-98,CatBriLio-01,CatBriLio-02,CanDelLew-08a}. If in addition one optimizes over
the positions of the nuclei, then crystallization is only known in 1D for Thomas-Fermi type models~\cite{BlaBri-02}.

\begin{figure}[ht!]
 \centering
 \includegraphics[width=12.5cm]{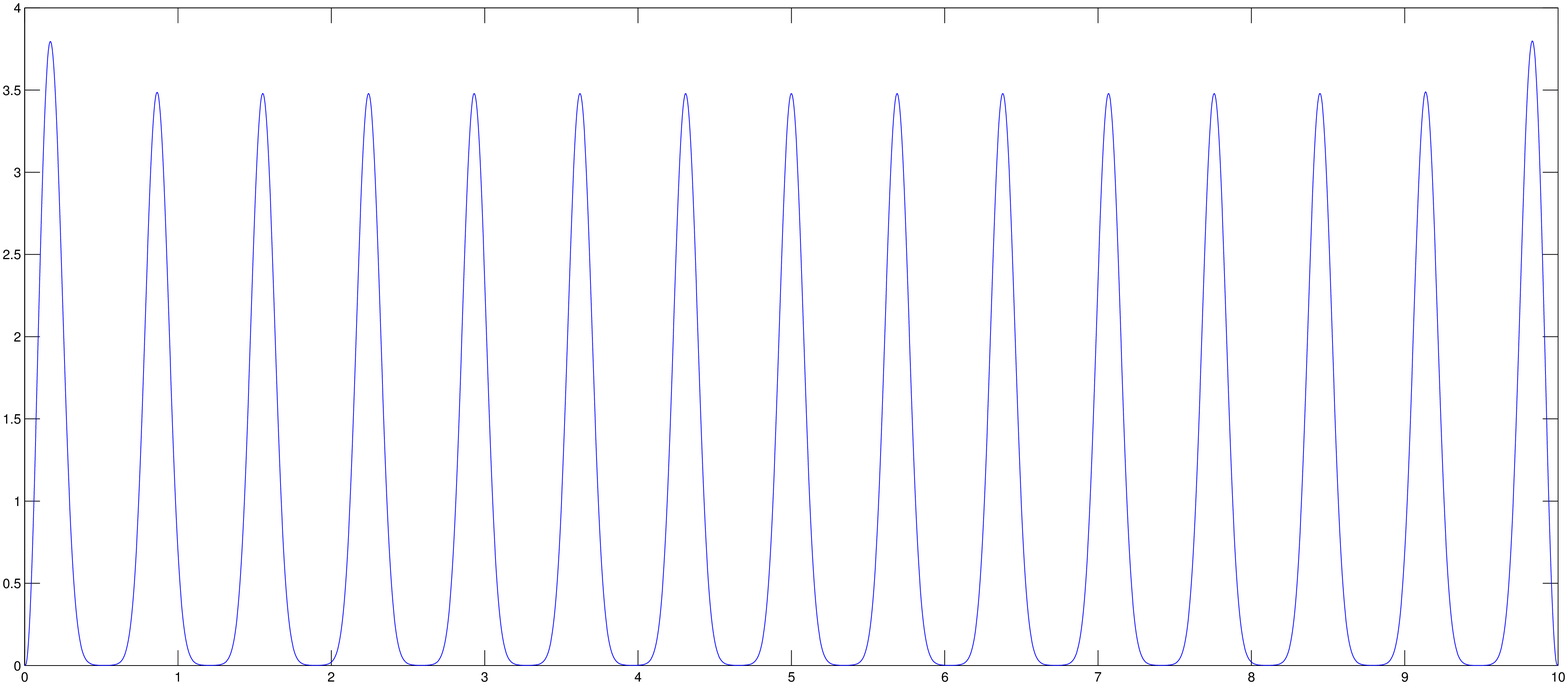}
\caption{Numerical calculation of the minimizing density $|u|^2$ of the Hartree energy~\eqref{eq:Hartree} in 1D for the truncated Lennard-Jones potential $w(x)=\min(10^3,|x|^{-12}-|x|^{-6})$. Here $N=10$, $\hbar^2/(2m)=1$, the average density is $\rho=1$ and we have used Dirichlet boundary conditions, on a mesh with $2\times10^3$ points on the interval $[0,10]$.\label{fig:Hartree}}
\end{figure}

For a general interaction, one could study the Hartree problem that is the mean-field approximation of the bosonic many-particle system~\cite{Lewin-15}. It relies on the nonlinear energy functional
\begin{multline}
\cE(u)=\frac{\hbar^2}{2m}\int_\Omega|\nabla u(x)|^2\,dx+\frac{1}{2}\int_\Omega \int_\Omega V(|x-y|)\,|u(x)|^2|u(y)|^2\,dx\,dy,\\
\text{with } \int_\Omega|u(x)|^2\,dx=N,
\label{eq:Hartree}
\end{multline}
which is similar in spirit to~\eqref{eq:regularized_Ohta-Kawasaki}. The numerical simulation presented in Figure~\ref{fig:Hartree} shows the occurrence of crystallization for this model in 1D for a Lennard-Jones interaction. Similar results have been observed for the potential $V(r)=\1_{(0,1)}(r)$ in higher dimensions~\cite{PomRic-94,JosPomRic-07,JosPomRic-07b,AftBlaJer-07,AftBlaJer-09,WatBra-12,KunKat-12,HenCinJaiPupPoh-12,AncRosToi-13}.

\subsection{Discrete systems}

In our review, we focused on continuous systems, defined in the whole space or in a domain $\Omega_N$ which grows as
$N\to+\infty$. An important literature is devoted to the study of \emph{discrete systems}. Such systems are defined
on a lattice $G\subset\R^d$, without assuming \emph{a priori} that the corresponding states are $G$-periodic. We expect
that proving the occurrence of periodicity is slightly easier, since a natural periodic lattice is already present in the
definition of the system. Many rigorous results have been proved for this kind of problems, although important
questions are still unsolved. The models considered can be either quantum or classical.

Examples of such systems are the (classical or quantum) Heisenberg or Ising models. Two main regimes
are usually dealt with: the ferromagnetic one, in which spins tend to be aligned with each other, and
antiferromagnetic in which spins are preferably of alternate sign. In this latter case, crystallization gives a
periodic lattice which size is twice that of the original one.

In 1986, Kennedy and Lieb have considered two systems of this type. In~\cite{KenLie-86,KenLie-86b,Lieb-86}, they study
electrons on a lattice, submitted to a pointwise interaction with fixed particles of opposite spin. They prove
that the electrons are located on a sub-lattice. In~\cite{KenLie-87} they consider a 1D system on the lattice
$\Z$. This model describes for instance deformations of a polyacetylene molecule. They prove that the minimizer is
periodic of period $2$, a phenomenon called \emph{Peierls instability}. This result has been further developed
in~\cite{LieNac-95,LieNac-95b,LieNac-95c}. It has been generalized in~\cite{GarSer-12}, and extended to the
hexagonal lattice in 2D in~\cite{FraLie-11}.

Apart from systems with analytical solutions, an important method for studying classical or quantum spin systems is
the \emph{reflection positivity} method. This strategy has been introduced in field theory~\cite{OstSch-73}, then
adapted and developed in the case of spin
systems~\cite{FroSimSpe-76,DysLieSim-78,FroLie-78,FroIsrLieSim-78,FroIsrLieSim-80}. This method aims at proving
phase transitions and long-range order. However, it does not always allow to conclude that the system is
periodic. For recent examples of application of this theory to crystallization problems, see for instance~\cite{GiuLebLie-06,GiuLebLie-07,GiuLieSei-13,GiuLieSei-14}.

\section*{Conclusion}

We have described several aspects of an important problem arising in physics and which, in spite of an intense activity, is
still not completely understood mathematically. In addition to the famous crystallization problem, several questions have been
mentioned, some of which are probably more at hand than others. Some progress in any of these directions would be of
high interest and would improve the theoretical understanding of the structure of matter at the microscopic scale. We hope that
this article will stimulate further research in these directions.

\bigskip

\noindent\textbf{Acknowledgement.} M.L. has received financial support from the European Research Council under the European Community's Seventh Framework Programme (FP7/2007-2013 Grant Agreement MNIQS 258023).


\end{document}